\documentclass[journal,a4paper,doublecolumn]{IEEEtran}

\usepackage[pdftex]{graphicx}
\usepackage{caption}
\usepackage{subfig}
\usepackage{color}
\usepackage[cmex10]{amsmath}
\usepackage{amssymb}
\usepackage{amsmath}
\usepackage{algorithmic}
\usepackage{algorithm}
\usepackage{tabularx}
\usepackage{soul}

\begin{document}
\title{A Sub-Nyquist Radar Prototype: Hardware and Algorithms}

\author{\IEEEauthorblockN{Eliahu Baransky\IEEEauthorrefmark{2}, Gal Itzhak\IEEEauthorrefmark{3}, Idan Shmuel, Noam Wagner, Eli Shoshan and Yonina C. Eldar\IEEEauthorrefmark{4}}
\IEEEauthorblockA{\\Department of Electrical Engineering\\Technion-Israel Institute of Technology, Technion City, Haifa, Israel\\
\IEEEauthorrefmark{2}selibar6@t2.technion.ac.il, \IEEEauthorrefmark{3}sgalitz@t2.technion.ac.il, \IEEEauthorrefmark{4}yonina@ee.technion.ac.il}}

\maketitle

\begin{abstract}
Traditional radar sensing typically employs matched filtering between the received signal and the shape of the transmitted pulse. Matched filtering is conventionally carried out digitally, after sampling the received analog signals. Here, principles from classic sampling theory are generally employed, requiring that the received signals be sampled at twice their baseband bandwidth. The resulting sampling rates necessary for correlation based radar systems become quite high, as growing demands for target distinction capability and spatial resolution stretch the bandwidth of the transmitted pulse. The large amounts of sampled data also necessitate vast memory capacity. In addition, real-time processing of the data typically results in high power consumption. Recently, new approaches for radar sensing and estimation were introduced, based on the Finite Rate of Innovation and Xampling frameworks. Exploiting the parametric nature of the radar problem, these techniques allow significant reduction in sampling rate, implying potential power savings, while maintaining the systems estimation capabilities at sufficiently high signal-to-noise ratios. Here we present for the first time a design and implementation of a Xampling based hardware prototype that allows sampling of radar signals at rates much lower than Nyquist. We demonstrate by real-time analog experiments that our system is able to maintain reasonable recovery capabilities, while sampling radar signals that require sampling at a rate of about $30\mbox{MHz}$ at a total rate of $1\mbox{MHz}$.
\end{abstract}
\begin{IEEEkeywords}
 radar, sampling, Compressed Sensing, Sub-Nyquist
\end{IEEEkeywords}

\IEEEpeerreviewmaketitle

\section{Introduction}
\label{sec:Introduction}
The classic radar sensing problem treats detection of targets moving in space. This is achieved by transmitting RF pulses of electromagnetic energy and sampling the signals caused by their reflection. The samples are then processed, in attempt to determine the targets$\text{'}$ location in space and their velocity. Traditional processing methods in both literature and practice involve a preliminary stage, referred to as matched filtering (MF) or pulse compression \cite{Cook,Richards}, in which the transmitted pulse is correlated with the received signal. For the problem of detecting targets in white Gaussian noise, the MF is known to maximize the effective signal-to-noise ratio (SNR).

The MF stage is typically carried out digitally, after having sampled the detected analog signal. Classic Shannon-Nyquist sampling theory \cite{Shannon} guarantees full recovery of a general bandlimited analog signal from samples taken at twice its baseband bandwidth. However, applying this framework to modern radar systems typically results in extremely high sampling rates, due to the wide bandwidth characteristic of the sampled signals. The latter is a direct consequence of the well-known relationship between a radar system's resolution and the bandwidth of the transmitted signals.

Formulating the radar problem as one of parametric inference, one may show that Nyquist sampling is, in fact, a redundant approach for resolving the desired parameters. Nonetheless, this method is still widely employed by modern radar systems, to a large extent because it produces a straightforward and simple solution to the parametric inference problem as well as to the preceding analog sampling stage. However, the growing sampling rates required due to the desire to increase resolution necessitate sophisticated analog front ends and imply higher power consumption and vast memory capacity.

Recently, new approaches \cite{Bajwa} to radar processing were introduced, which allow practical solution of the parametric problem, from a small number of measurements taken after appropriate analog pre-filtering. The estimation problem is solved using known tools from array processing \cite{StoicaMoses}, with the necessary number of measurements typically much smaller than that obtained by Nyquist sampling. Related works \cite{Tur,Gedal,Wagner} treat ultrasound signal sampling and are readily adapted to the radar scenario. These approaches are based on the Finite Rate of Innovation (FRI) \cite{Vetterli} and Xampling~\cite{Mishali,Mishali1} approaches. The FRI framework treats sampling and recovery of signals characterized by a finite number of degrees of freedom per unit time. In many cases such signals can be sampled and recovered at a rate proportional to the number of unknowns per time interval, which is usually much lower than the Nyquist rate. The Xampling philosophy ties together sub-Nyquist sampling based on analog pre-processing with techniques of compressed sensing (CS) \cite{CandesCSIntro,Eldar,Duarte} for recovery. However, as discussed in the following sections, these approaches typically require sophisticated sampling schemes, which acquire generalized measurements of the analog signals. Instead, we present a concrete analog to digital conversion (ADC) scheme and a recovery algorithm with relaxed constraints, which achieves similar performance. We further introduce an analog hardware prototype which, implementing our proposed prototype, allows reconstruction  of radar signals from low-rate samples which carry sufficient information about the parameters of interest.

Our hardware prototype implements a combination of the multi-channel topology suggested in \cite{Gedal} and the filtering approach presented in \cite{Tur}, additionally taking into account the practical challenges they impose. In particular, our board consists of 4 channels, each comprising a bandpass crystal filter with a random effective carrier frequency. This allows to obtain a wide spread of Fourier coefficients of the signal in an efficient manner. The proposed recovery algorithm then uses these coefficients to recover the delays and amplitudes of the radar signal. Using crystal filters, which have extremely narrow transition bands, we are able to obtain sufficient amount of information from the signal, while substantially decreasing the total sampling rate, as discussed in Section~\ref{sec:Hardware}. Since crystal filters are standard, of-the-shelf components, we are confined to adapt our channel design to their properties, in order to maximize their efficiency. We discuss the challenges this imposes and our method for overcoming them in Section~\ref{sec:Hardware}.

Simulations as well as real-time experiments of our hardware prototype prove high target hit-rate and location estimation capabilities for pre-integration SNR values over -14dB, with a large reduction factor in sampling rate compared to Nyquist sampling. In fact, standard radar systems, in which Nyquist sampling is employed for solving the parametric inference problem, typically oversample the received signals, due to the non-ideal behavior of practical anti-aliasing filters. With respect to the resulting sampling rates, we have been able to achieve a thirty-fold reduction using our approach. That is, our system operates at a total sampling rate of $1\mbox{MHz}$ (20 times less than the signal's Nyquist rate), while an MF based system would typically operate at a rate of $30\mbox{MHz}$ (requiring an over-sampling factor of 1.5 in order to use practical filters). We are able to achieve this reduction without substantially degrading the target hit-rate and location estimation capabilities, provided that the system operates at sufficiently high SNR. For instance, in our setup we achieved target hit-rate of approximately 90\% for pre-integration SNR's larger than -12 dB.

To estimate the underlying sparse structure from the selected Fourier coefficients of the received signal, or the targets in our setting, previous work in the FRI framework traditionally employed spectral estimation techniques. Here, we use a CS formulation of the recovery stage, as suggested in \cite{Wagner}. 
There have been several others works that employ CS algorithms in the context of radar such as \cite{Herman,BlockSparse,CompressiveRadarImaging,AdaptiveCSRadar}.
However, these papers either do not address sample rate reduction and continue sampling at the Nyquist rate, or they assume that a smaller number of samples has been obtained however they do not address how to sub-sample the analog received signal using a practical ADC. Furthermore, previous CS-based methods typically impose constraints on the radar transmitter.

In CS, the signal is assumed to have a sparse representation in a discrete basis.  An extensive search for the best sparse representation results in combinatorial run-time complexity, which is impractical for real-time applications such as radar sensing. Many polynomial-time algorithms have been proposed that can be shown to recover the true sparse vector under appropriate conditions \cite{Eldar}. In our simulations, we use the orthogonal matching pursuit (OMP) algorithm \cite{OMP,Tropp}.

The remainder of the paper is organized as follows. Section~\ref{sec:Xampling_of_Radar_Signals} establishes the mathematical foundation of our techniques. We present our radar signal model, link several CS concepts to our application and provide justification for our chosen parameters. Section~\ref{sec:Hardware} compares different analog sampling implementations. These systems offer a compromise between theoretical requirements and practical hardware constraints. We then elaborate on the design of our hardware. Finally, in Section~\ref{sec:sims_exp} we present results of MATLAB hardware simulations, as well as the results of real-time hardware experiments. We also describe our realization environment, and conclude by discussing the overall system performance.

\section{Xampling of Radar Signals}
\label{sec:Xampling_of_Radar_Signals}
\subsection{Methodology}
Radar systems estimate target locations by transmitting periodic streams of pulses and processing their reflections. We model the received radar signal as the following stream of pulses:
\begin{equation}
x(t)=\sum_{n=0}^{M-1} \sum_{l=1}^{L} a_l h(t-nT-t_l),~a_l \in \mathbb{C},~t_l \in \left[0,T\right),
\label{eq:Model}
\end{equation}
where $T$ is the radar's pulse repetition interval (PRI) and $M$ is the number of transmitted pulses. This model complies with monostatic radars, with non-fluctuating point targets assumed to be stationary or moving at very slow velocities, so that no Doppler shift is incorporated into the model. We show how to adapt the results to allow for Doppler processing in \cite{BE13}. The parameters $\{a_l,t_l\}_{l=1}^{L}$ correspond to the estimated pulses' amplitudes and  delays respectively, and are proportional to the targets' distance from the receiver and their radar cross section (RCS). We assume that the shape of the pulse $h(t)$ and the maximal number of echoes $L$ are known, although future research may relax this constraint. In particular, knowledge of $L$  is used to simplify the stopping criteria of our reconstruction algorithm which will be presented in the next part, but is not essential.

Traditional radar systems sample the received signal at the Nyquist rate, determined by $h\left(t\right)$'s baseband bandwidth.  Our goal is to recover $x\left(t\right)$ from its samples taken far below this rate.  It is readily seen that $x\left(t\right)$ is completely defined by at most $2L$ unknown parameters, namely $a_l$ and $t_l$, within any length-$T$ time interval. Hence, in the absence of noise, one would expect to be able to accurately recover $x\left(t\right)$ from only $2L$ samples per time $T$ \cite{Vetterli}.  If $L/T$ is sufficiently small with respect to $h\left(t\right)$'s bandwidth, then this implies a significant reduction in sampling rate. Since radar signals tend to be sparse in the time domain, simply acquiring $2L$ data samples would not yield adequate recovery, as most samples would not contain any information with high probability. This implies that a pre-sampling analog processing must be performed in order to smear the signal in time. 

As shown in \cite{Vetterli,Tur,Gedal}, the $2L$ unknowns defining $x(t)$ may indeed be recovered from only $2L$ measurements, corresponding to $x(t)$'s projection onto a subset of its Fourier series coefficients. Calculated with respect to a single period $\left[0,T\right)$, these coefficients are written as
\begin{equation}
\begin{split}
X[k]&   =\frac{1}{T}\int_{-\infty}^{\infty}\left[\sum_{l=1}^{L}a_l h(t-t_l)\right]e^{-j\frac{2\pi}{T}kt}dt \\
   &= \frac{1}{T} \sum_{l=1}^{L} a_l \int_{-\infty}^{\infty}h(t-t_l) e^{-j\frac{2\pi}{T}kt}dt \\
   &= \frac{1}{T} H\left(\frac{2\pi}{T}k\right) \sum_{l=1}^{L} a_l e^{-j\frac{2\pi}{T}kt_l},
\end{split}
\label{eq:FourierSeries}
\end{equation}
where $H(\omega)$ is the continuous-time Fourier transform (CTFT) of the pulse. Choosing the coefficients such that $H\left(\frac{2\pi}{T}k\right)$ is nonzero, \eqref{eq:FourierSeries} can be rewritten as
\begin{equation}
Y[k]=\frac{X[k]}{\frac{1}{T}H\left(\frac{2\pi}{T}k\right)}= \sum_{l=1}^{L} a_l e^{-j\frac{2\pi}{T}kt_l},
\label{eq:Measurements}
\end{equation}
which is a standard sum-of-exponentials problem. It can be shown that this problem has a unique solution given $K \geq 2L$ coefficients \cite{Vetterli}. We discuss the hardware for the acquisition of these measurements in Section~\ref{sec:Hardware}. The rest of this section is devoted to reviewing algorithmic approaches for solving \eqref{eq:Measurements}.

\subsection{Recovery Algorithm}
Many mature techniques for solving \eqref{eq:Measurements} exist, among which are matrix pencil \cite{Sarkar}, annihilating filter \cite{Vetterli} and many others that can be found in \cite{StoicaMoses}. These techniques arise from spectral analysis frameworks, and generally require the set of measurements to form a consecutive subset of the signal's Fourier coefficients. An exception is the MUSIC algorithm which can be applied on any set of coefficients \cite{MUSIC}. While these methods work well at high SNR, their performance deteriorates at low SNR values.

In \cite{Wagner} it was suggested to use a non-consecutive set of Fourier coefficients selected in a distributed manner, as many detection systems (namely, ultrasound in \cite{Wagner} and radar in our work) benefit from wide frequency aperture. While consecutive coefficients can be obtained using a simple low-pass filter, it is shown in \cite{Wagner} that a distributed selection results in better recovery and noise robustness. In order to remove the constraint of consecutive selection, and improve performance in low SNR, we choose to employ a CS formulation of \eqref{eq:Measurements} for recovery.

We begin by quantizing the analog time axis with a resolution step of $\Delta$, thus, approximating \eqref{eq:Measurements} as
\begin{equation}
Y[k] \approx \sum_{l=1}^{L} a_l e^{-j\frac{2\pi}{T}kn_l\Delta}, \quad  0 \leq n_l < N,
\label{eq:TimeQuant}
\end{equation}
where $N=T/\Delta$ is the number of bins in the PRI and $t_l \approx n_l\Delta$ is the discrete approximation of the time delays. Selecting a finite subset of $K$ measurements, $\kappa=\{k_1,k_2,\ldots,k_K\} \subset \{0,\ldots,N-1\}$, \eqref{eq:TimeQuant} may be written as
\begin{equation}
\mathbf{y}=\mathbf{A}\mathbf{x}.
\label{eq:MatForm}
\end{equation}
Here $\mathbf{A}$ is a $K \times N$ matrix, formed by taking the set $\kappa$ of rows from an $N\times N$ DFT matrix, and $\mathbf{x}\in\mathbb{C} ^ N$ is an $L$-sparse vector with nonzero entries at indices $\{n_l\}_{l=1}^{L}$. In  the context of CS, $\mathbf{A}$ is known as the sensing matrix.

Our goal is to find the nonzero entries of $\mathbf{x}$ from the measurements $\mathbf{y}$.  This is a standard CS problem, with $\mathbf{A}$ being a partial Fourier operator. A solution can be obtained, for example, by using the well-known OMP algorithm. The algorithm iteratively finds the nonzero entries of $\mathbf{x}$ by seeking the maximal correlations between $\mathbf{y}$ and the columns of $\mathbf{A}$, while maintaining an orthogonalization step at the end of each iteration \cite{Tropp,OMP}. To further improve performance, we used a variation of the standard OMP algorithm where we computed the maximal correlation using the pseudo-inverse rather than the inner product. The pseudo-inverse, denoted by $(\cdot)^\dagger$, is defined as:  $A^{\dagger}=(A^{H}A)^{-1}A^{H}$.

Once the nonzero entries are found, the time delays are directly calculated and the amplitudes are estimated via a standard least-squares technique. A pseudo-code of the recovery algorithm is given in Algorithm~\ref{alg:OMP}.
\begin{algorithm}[h]
\caption{OMP Algorithm}
\begin{algorithmic}
\REQUIRE Measurement vector $\mathbf{y} \in \mathbb{R}^K$; Sensing matrix $\mathbf{A}\in\mathbb{R}^{K\times N}$; Quantization step $\Delta$;
\ENSURE Estimated time delays $\{\hat{t}_{l}\}_{l=1}^{L}$; Corresponding estimated amplitudes $\{\hat{a}_{l}\}_{l=1}^{L}$
\STATE $\mathbf{r}\leftarrow\mathbf{y}$; $\Omega\leftarrow\phi$ \COMMENT{Initialization}
\FOR{$l=1$ to $L$}
\STATE $\mathbf{p}=\mathbf{A}^{\dagger}\mathbf{r}$ \COMMENT{Acquire subspace approximation}
\STATE $\hat{n}_{l}=\underset{n=1,\ldots,N}{\arg\max} |p_{n}|$ \COMMENT{Find maximal component}
\STATE $\Omega\leftarrow\Omega\cup\{\hat{n}_{l}\}$ \COMMENT{Augment the set of selected columns' indices}
\STATE $\hat{t}_l=\hat{n}_{l}\Delta$
\STATE $\mathbf{A}_{\Omega} \leftarrow \left[ \mathbf{A}_{\hat{n}_{1}} \cdots \mathbf{A}_{\hat{n}_{l}} \right]; (\mathbf{A}_{n})_{m}=\mathbf{A}_{m,n}, m=1,\ldots,K;$
\STATE $\mathbf{P}_{\Omega}=\mathbf{I}-\mathbf{A}_{\Omega}\mathbf{A}^{\dagger}_{\Omega}$ \COMMENT{Subspace projection operator}
\STATE $\mathbf{r}=\mathbf{P}_{\Omega}\mathbf{y}$ \COMMENT{Compute the residual}
\ENDFOR
\STATE $\mathbf{a}=\mathbf{A}^{\dagger}_{\Omega}\mathbf{y}$ \COMMENT{Estimate amplitudes via least-squares}
\end{algorithmic}
\label{alg:OMP}
\end{algorithm}

\subsection{Frequency Selection}
As mentioned previously, CS based techniques allow flexibility in choosing the Fourier coefficients. Using OMP, high recovery performance is promised, provided that the sensing matrix satisfies desired properties such as the restricted isometry property (RIP).  Selecting the frequency samples uniformly at random, it is known that if
\begin{equation}
K \geq CL\left(\log{N}\right)^{4},
\label{eq:bound}
\end{equation}
for some positive constant $C$, then $\mathbf{A}$ obeys the RIP with high probability \cite{Vershynin}. In contrast, for consecutive selection the RIP is not generally satisfied, unless the cardinality of $\kappa$ is significantly increased. However, applying random frequency sampling is not practical from a hardware perspective, and therefore a rule for selecting a good constellation of frequency samples is desired.
\begin{figure}[htb]
\centering
\includegraphics[width=0.5\textwidth]{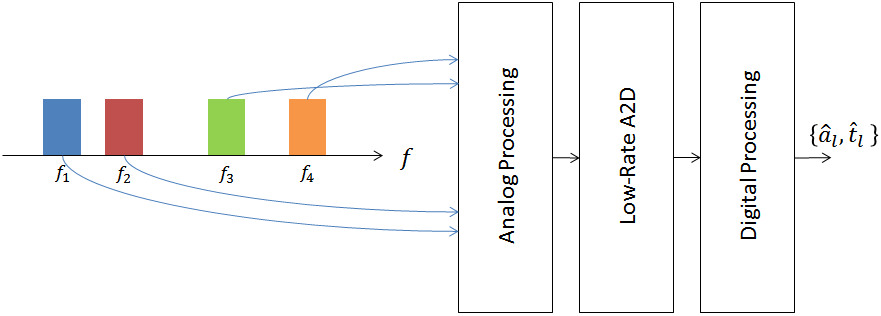}
\caption{Xampling of radar signals.}
\label{fig:Xampling}
\end{figure}
Some practical guidelines for choosing the frequencies are  suggested in \cite{Stoica}. The authors consider an equivalent problem to \eqref{eq:Measurements}, with the roles of frequency and time interchanged. That is, the measurement vector consists of time samples of the signal, which obey a spectral line model. Conclusions can be easily adapted to our application with minor changes. Applying these guidelines, we may formalize the following relationship between the support of the recovered signal $T$, the grid resolution $\Delta$, and the chosen frequency samples $\{f_{i}\}_{i=1}^{K}$:
\begin{equation}
T=\frac{1}{\underset{i,j=1,\ldots,K;~i\neq j}{\text{min}}\left|f_i-f_j\right|},
\label{eq:StoicaPRI}
\end{equation}
\begin{equation}
\Delta=\frac{1}{c \cdot \underset{i,j=1,\ldots,K}{\text{max}}\left|f_i-f_j\right|},
\label{eq:StoicaGrid}
\end{equation}
where $c$ is an empirical constant. In our simulations we set $c=20$, finding this value to be more robust than that proposed in \cite{Stoica}.

Condition \eqref{eq:StoicaPRI} constraints us to choosing at least two Fourier coefficients consecutively. In our application, we choose a constellation consisting of four groups of coefficients,  where the coefficients in each group are consecutive.   We refer to this choice as multiple band-pass sampling.  Note, that \eqref{eq:StoicaPRI} could be equally satisfied by simply choosing a single group of $K$ consecutive coefficients.  As discussed in Section~\ref{sec:Hardware}, such a choice can be implemented by simple hardware, employing a single low-pass filtering channel.  However, our multiple band-pass constellation has the advantage of acquiring the  measurements over a wider frequency aperture.  At the same time, it still allows practical hardware implementation (as detailed in Section~\ref{sec:Hardware}). Referring to \eqref{eq:StoicaGrid}, by widening the frequency aperture we may employ a finer grid resolution during the recovery process.  Moreover, empirical results show that highly distributed frequency samples provide better noise robustness~\cite{Wagner}.  We point out that widening the frequency aperture eventually requires increasing the number of samples $K$, otherwise recovery performance may degrade.  This trade-off, observed in our experiments, is readily seen from \eqref{eq:bound}, where the minimal number of samples $K$ gradually increases with $N$, corresponding to the grid resolution.

An intuitive explanation for our choice of coefficients may be obtained by adopting insights from the field of array processing. This is because our sampling domain (Fourier coefficients) is related to the parameters' domain (time delays) in the exact same manner in which the geometric sensor deployment pattern in array processing is related to the resulting beampattern. The beampattern determines the angular resolution and ambiguity of the array, which are analogous to temporal resolution and ambiguity in our problem. In particular, narrower array aperture (in our problem, span of Fourier coefficients) will typically result in a wider main-lobe, providing poorer target resolution, as well as degradation in the accuracy of direction of arrival estimation (DOA) at lower SNR. Considering an equally spaced sensor array, one can show that widening the array aperture while maintaining angular ambiguity requires additional sensors; trying to distribute the sensors further apart while maintaining their uniform distribution would inevitably affect directional ambiguity. A possible trade-off is to distribute the sensors non-uniformly within the wide aperture, for instance in a pattern chosen randomly, such that the minimal distance between at least two sensors remains fixed. 

\begin{figure*}[!ht]
\centering
\subfloat[]{\includegraphics[width=0.53\textwidth]{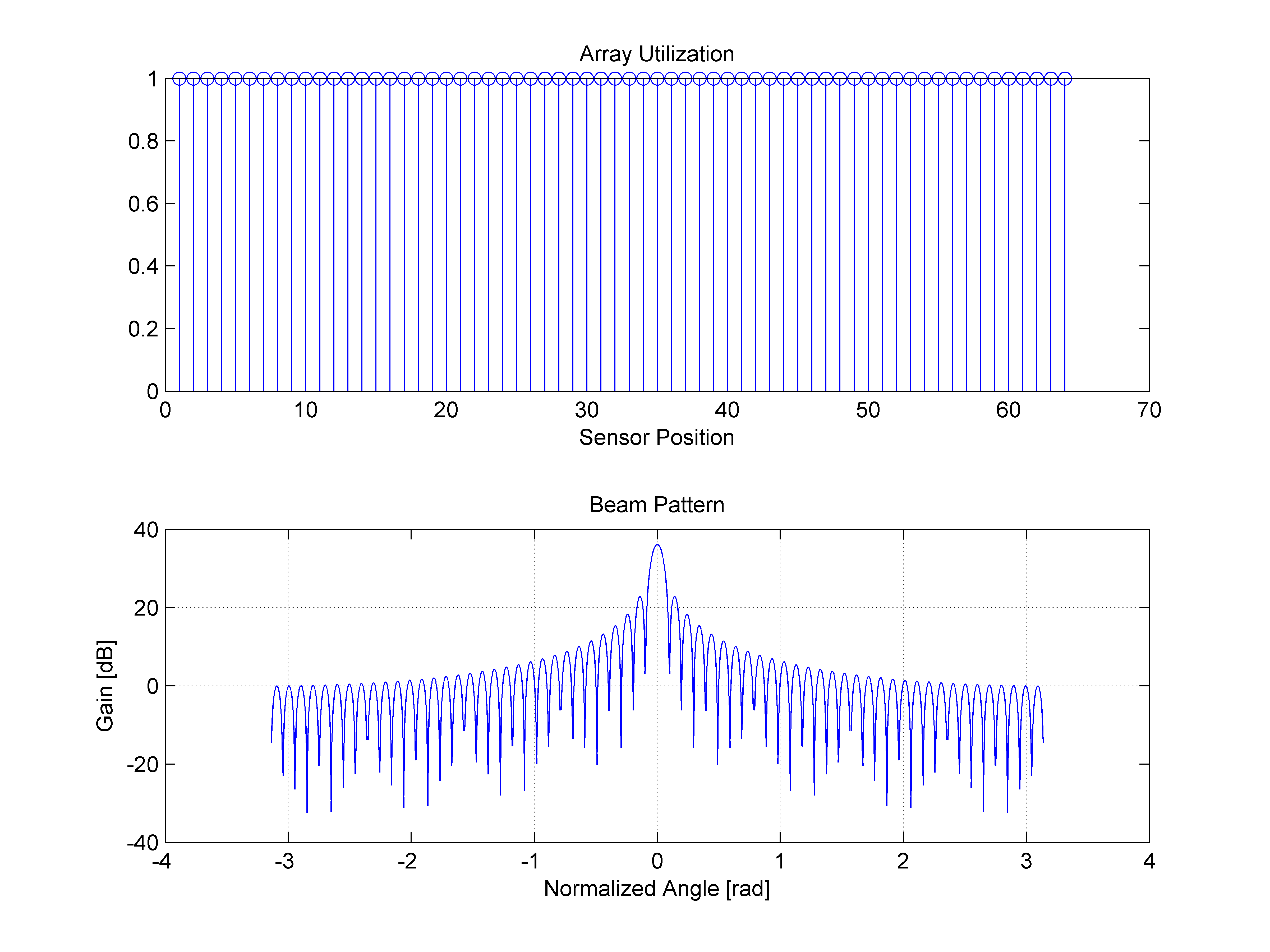}}  \subfloat[]{\includegraphics[width=0.53\textwidth]{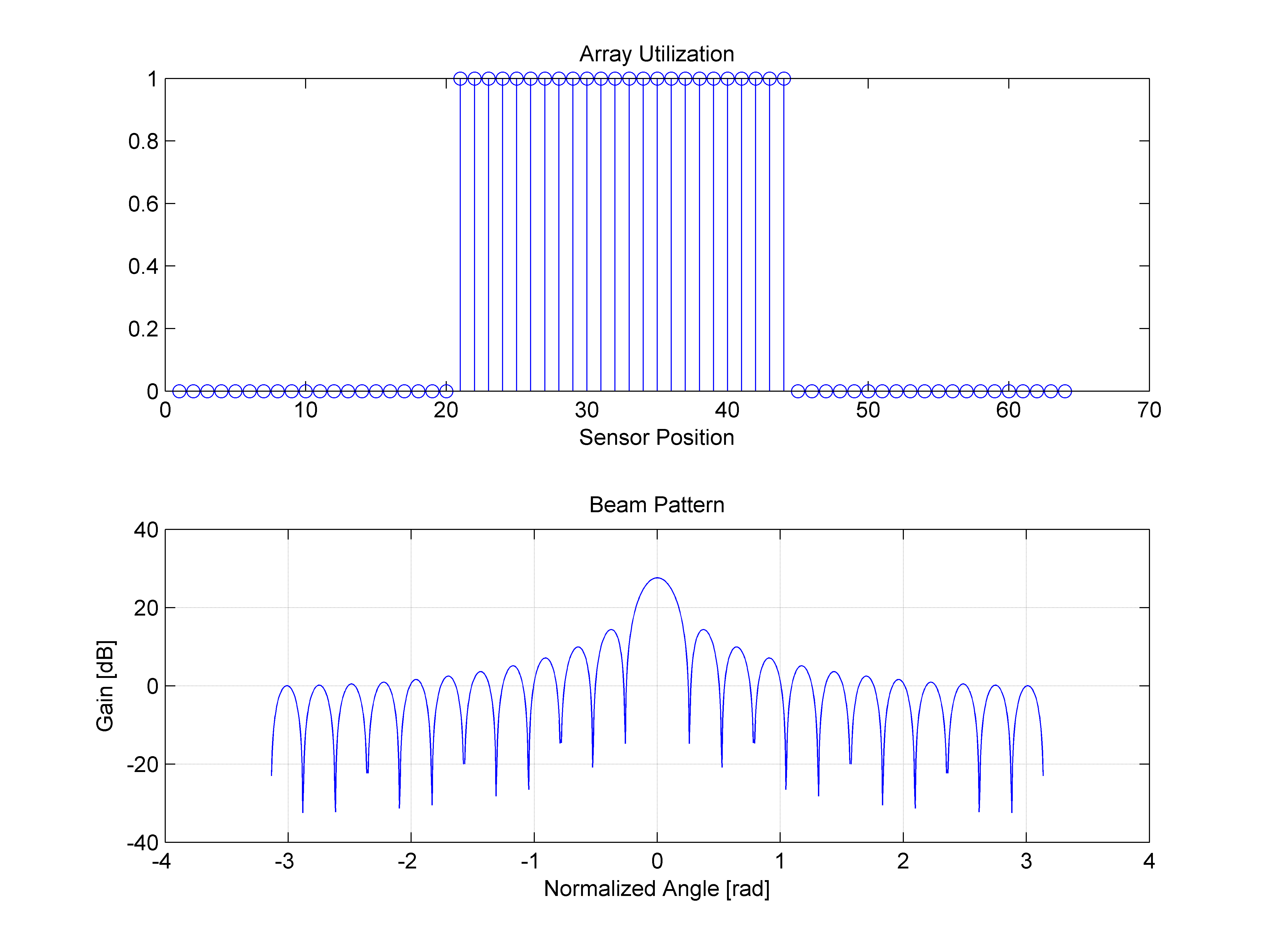}}  \\
\subfloat[]{\includegraphics[width=0.53\textwidth]{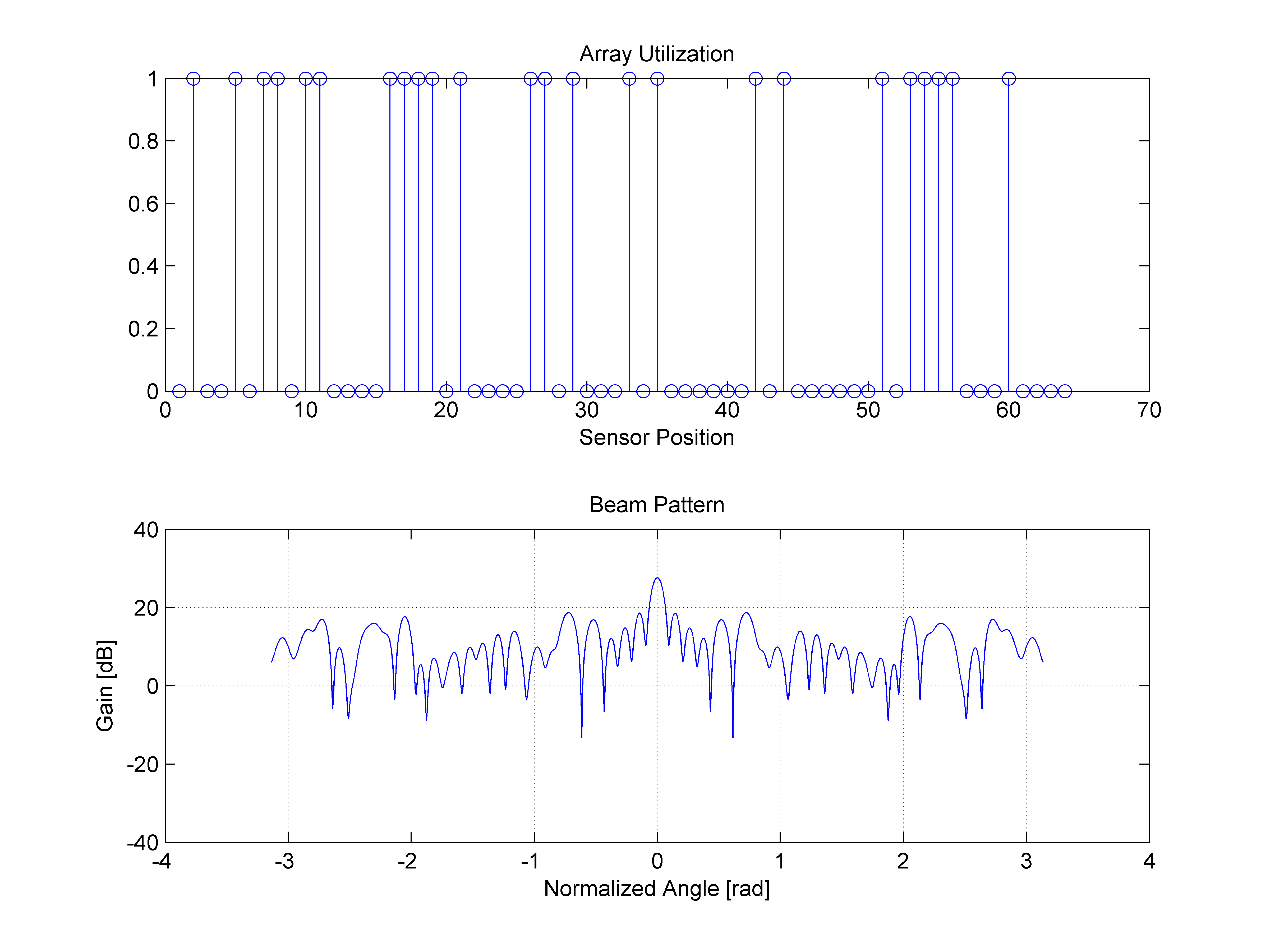}}  \subfloat[]{\includegraphics[width=0.53\textwidth]{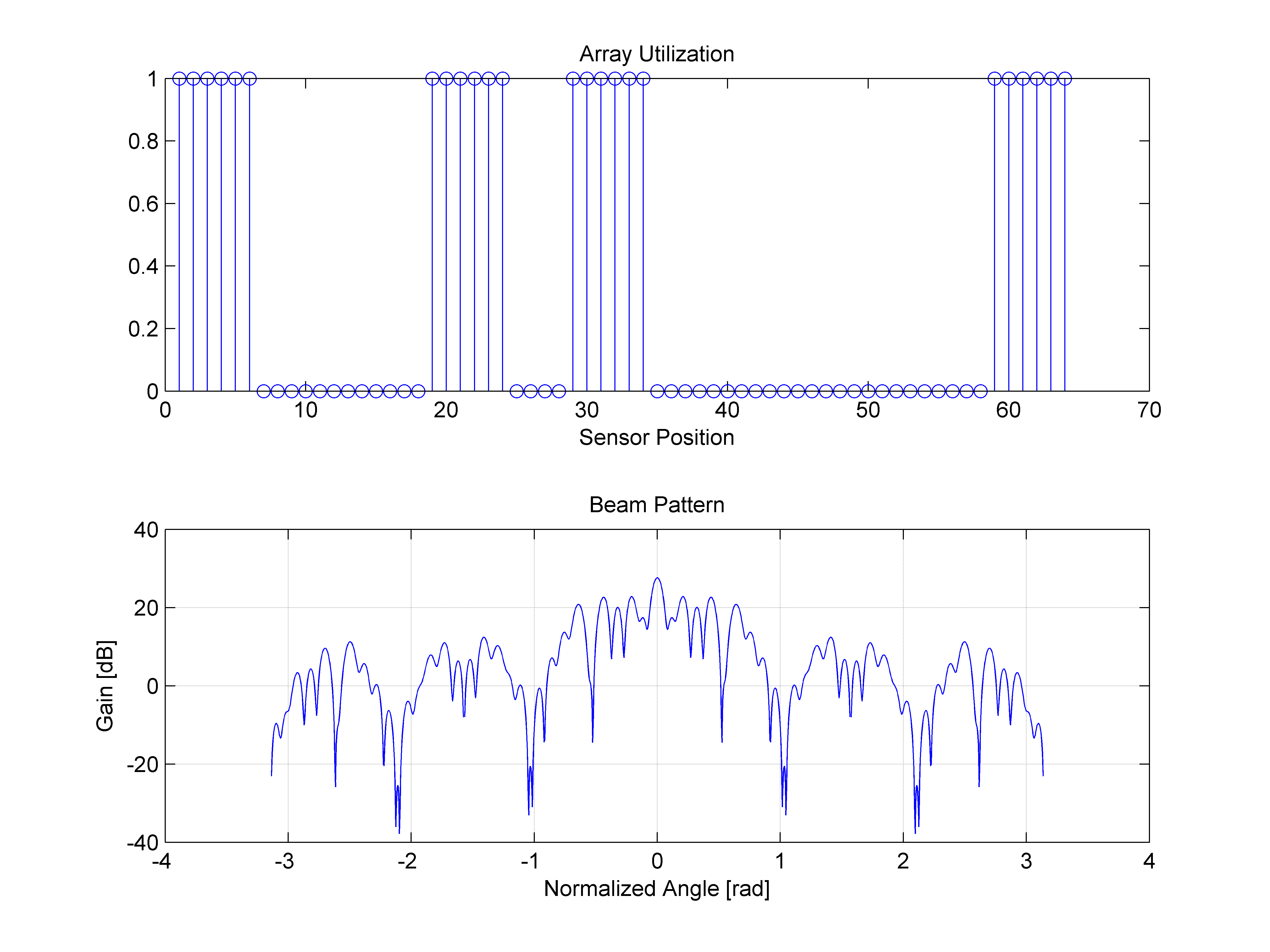}}
\caption{Beam patterns for different geometries of sensor arrays. (a) Full array. (b) Consecutive. (c) Random. (d) Four distributed groups.}
\label{fig:arrayproc}\end{figure*}

In Fig.~\ref{fig:arrayproc} we illustrate four choices of sensor patterns, together with the resulting beampatterns. We begin with a uniform array comprising 64 sensors. We then look at three arrays, each employing only $24$ sensors: consecutive array where the sensors are bunched together, a randomly chosen array, and an array comprising $4$ groups of size $6$ each where the groups are randomly distributed. The latter constellation resembles our choice of Fourier coefficients.
Notice how the non-uniform, randomly chosen distribution of sensors, maintains a narrow mainlobe, at the cost of increased sidelobes. The latter may result in aliasing of sufficiently strong targets. The random groups maintain the narrow mainlobe with a slight increase in the sidelobes. From a hardware perspective, this selection is far more practical than a completely random choice. Thus, it offers a reasonable trade-off between performance and hardware design.
\section{Hardware}
\label{sec:Hardware}
We now present practical considerations which guided us through the design of our analog board. We begin by examining previous proposals for sub-Nyquist schemes and point out the difficulty in their direct implementation.  We then present our solution. Throughout this section we consider the following system parameters: we are working with a $T=1\mbox{msec}$ PRI, which corresponds to $1\mbox{KHz}$ spacing of Fourier coefficients. Our pulse is approximately flat in spectrum, over the extent of $10\mbox{MHz}$ (single-sided band). We assume a maximal number of $L=6$ targets within the PRI.

\subsection{Previous Work}
Extracting a consecutive Fourier subset can be performed using a LPF, followed by sampling at twice its stop frequency. The DFT of the samples provide the desired Fourier coefficients. However, as discussed in Section~\ref{sec:Xampling_of_Radar_Signals}, recovery performance is enhanced when using a set of coefficients distributed over a larger part of the signal's spectrum.

For the extraction of arbitrary sets of Fourier coefficients, \cite{Tur} introduces a single-channel, pre-sampling filtering process, with the following filter:
\begin{equation}
G(f)=\left\{
  \begin{array} {l l}
  \text{nonzero}, & \quad f=\frac{k}{T}, k \in \kappa \\
  0, & \quad f=\frac{k}{T}, k \in \mathbb{Z} \setminus \kappa \\
  \text{arbitrary}, & \quad \mbox{otherwise}.
  \end{array} \right.
\label{eq:SoSFilter}
\end{equation}
With an arbitrary choice of isolated frequency samples, a kernel satisfying \eqref{eq:SoSFilter} would be characterized by multiple pass-bands and extremely high frequency selectivity. In practice, these characteristics are difficult to satisfy when designing a practical analog filter.

The selectivity property requires high Q-factor filters, characterized by a large attenuation within $1\mbox{KHz}$, the coefficients' spacing. Filtering under this regime requires Q-factors in the order of thousands, which is infeasible even when using piezoelectric components. Aside from the problem of achieving the high Q-factor, any off-the shelf filter with high Q-factor will have a long impulse response. This will cause instability when processing multiple pulse-streams. Finally, there are no off-the-shelf multiple pass-band components. These considerations show that implementing this class of filters with our application's specification is difficult.

Another work \cite{Gedal} suggests the use of multi-channel mixers and integrators to directly compute and sample the Fourier coefficients at a rate of $1/T$ in each channel. To obtain complex-valued Fourier coefficients, I/Q channels must be employed. This doubles the number of channels, and requires accurate synchronization between the I/Q demodulators. Furthermore, implementation of multi-channel circuits in hardware results in a complicated system, characterized by large physical dimensions and high number of components, requiring synchronization.

\begin{figure*}[htb]
\includegraphics[width=1\textwidth]{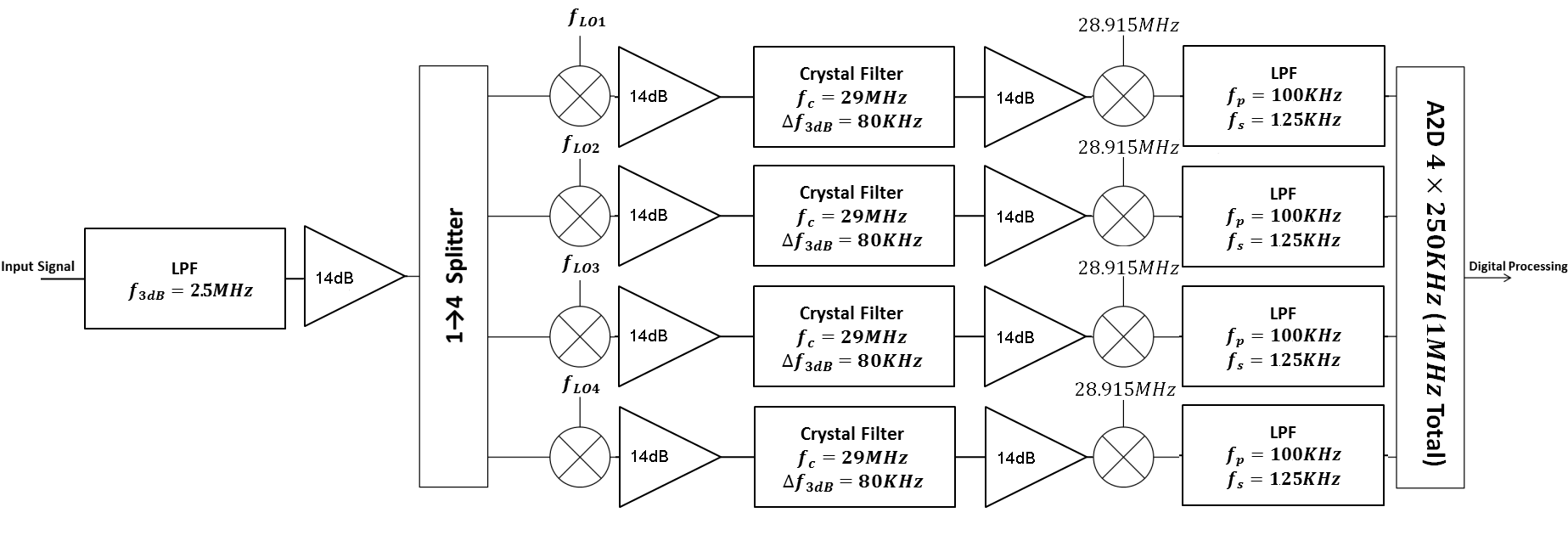}
\caption{A block diagram of the 4-channel crystal receiver. The four up-modulating LO's are as follows: ${f}_{LO1}$= $28.375 \mbox{MHz}$,  ${f}_{LO2}$=$28.275 \mbox{MHz}$, ${f}_{LO3}$= $27.65 \mbox{MHz}$,  ${f}_{LO4}$=$27.391 \mbox{MHz}$. }
\label{fig:Scheme}
\end{figure*}

\subsection{Multi-Channel Crystal Receiver}

Here we present a multi-channel crystal receiver approach to obtain Fourier coefficients in a manner that is both practical and efficient. This approach makes use of four parallel channels which sample distinct bands of the radar signal spectrum, as illustrated in Fig.~\ref{fig:Xampling}. This is achieved through filtering the desired band, demodulating it to baseband and then sampling it at its Nyquist rate. In this scheme, instead of sampling isolated Fourier coefficients, we acquire four independent sets of consecutive harmonies. This is a trade-off between the theoretical algorithmic requirements, which would benefit from a fully distributed selection, and the constraints of practical analog filters.  

In our scenario,  the proposed approach achieves a $1\mbox{MHz}$ sampling rate (combining all four channels), whereas the Nyquist rate, corresponding to $h\left(t\right)$'s bandwidth, is $20\mbox{MHz}$. However, the design of systems for classical methods such as MF, is constrained to the usage of practical filters as well. An anti-aliasing LPF must be used prior to sampling, and due to the finite width of its transition band, it is not feasible to sample the filter's output at the signal Nyquist's rate without causing aliasing. Modeling a LPF with Chebyshev type-I, and allowing a maximal order of $6$, we were able to achieve a stop frequency of $15\mbox{MHz}$, which requires sampling at $30\mbox{MHz}$, a 1.5 oversampling factor with respect to the signal's Nyquist rate. Thus we conclude that our scheme achieves an even greater reduction compared to practical implementations of classical methods.

In order to maintain a low oversampling factor, we employ filters characterized by narrow pass bands.  For instance, with $T=1\mbox{msec}$, a $120\mbox{KHz}$ pass band corresponds to 120 Fourier coefficients.  This is ten times the minimal number implied by the FRI framework for $L=6$ pulses in a noiseless scenario.  In order to avoid the usage of I/Q channels which, as stated before, adds complexity to the system, we demodulate the BPF's stop frequency in each channel, rather than its central frequency, as done in customary design.  This requires that the filters are characterized by narrow transition bands, in order to sufficiently attenuate image frequencies aliasing our chosen coefficients. We observed that active filters satisfying our narrow pass-band requirement yield unsatisfactory attenuation of the image frequencies, degrading the performance of the recovery algorithm. We hence chose to use crystal filters, whose transition bands are extremely narrow. These filters are characterized by a $80\mbox{KHz}$ pass band.  The narrow transition bands allow us to achieve low-rate sampling, while extracting a sufficient number of Fourier coefficients. Another advantage of the crystal filtering implementation is that it results in identical receiver channels, which have small phase unbalance and are easy to synchronize.

We performed simulations in order to obtain a good combination of the $80\mbox{KHz}$ pass bands in all four channels: multiple constellations of four frequency groups were examined, each obtained by randomly choosing four central frequencies and then taking narrow sets of consecutive coefficients surrounding these frequencies.  The following constellation was found to yield good performance, based on the evaluation methods detailed in Section~\ref{sec:sims_exp}: 590KHz - 670KHz; 690KHz -770KHz; 1315KHz - 1395KHz; 1574KHz -1654KHz.
\begin{figure*} [htb!]
\centering
\subfloat[]{\includegraphics[width=0.35\textwidth]{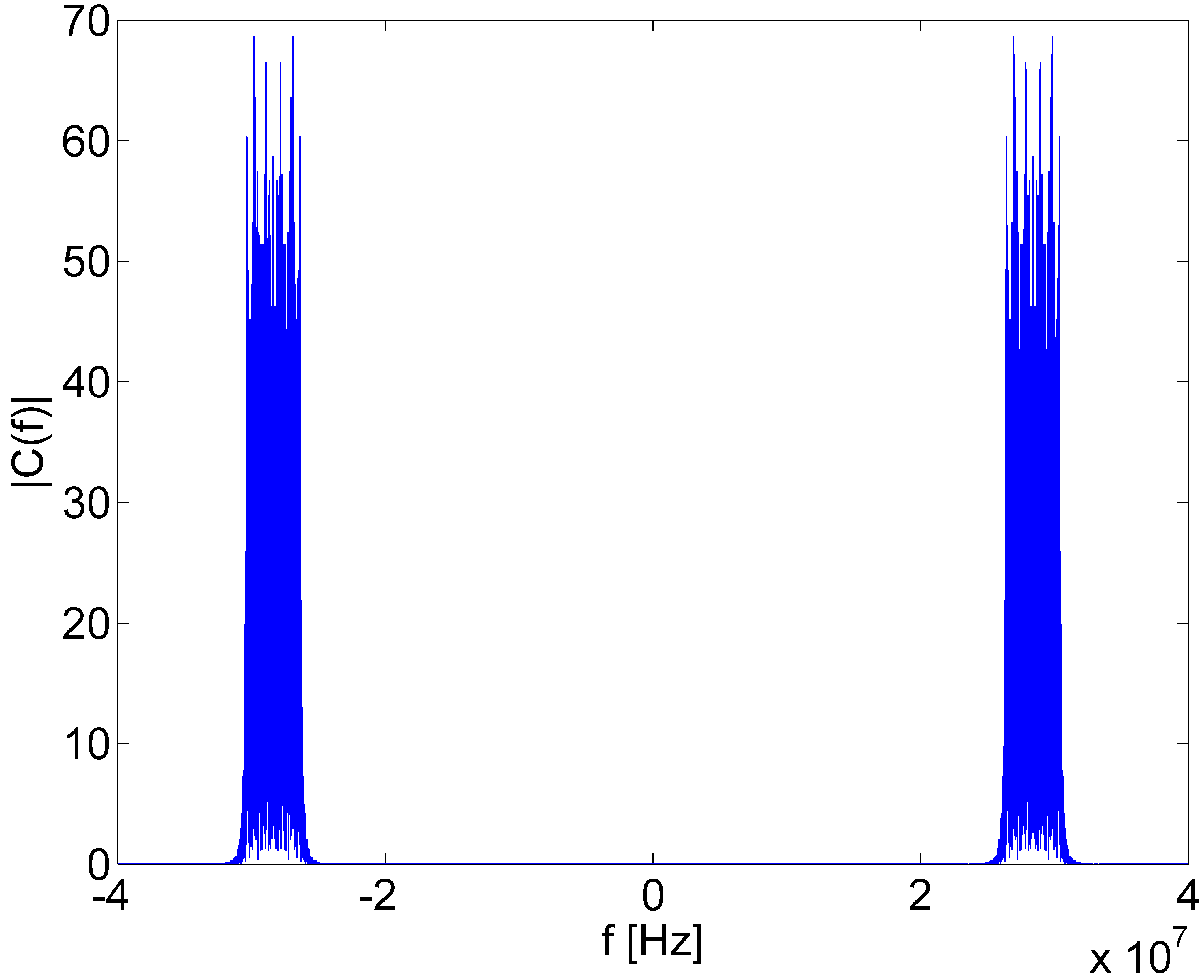}}  \subfloat[]{\includegraphics[width=0.35\textwidth]{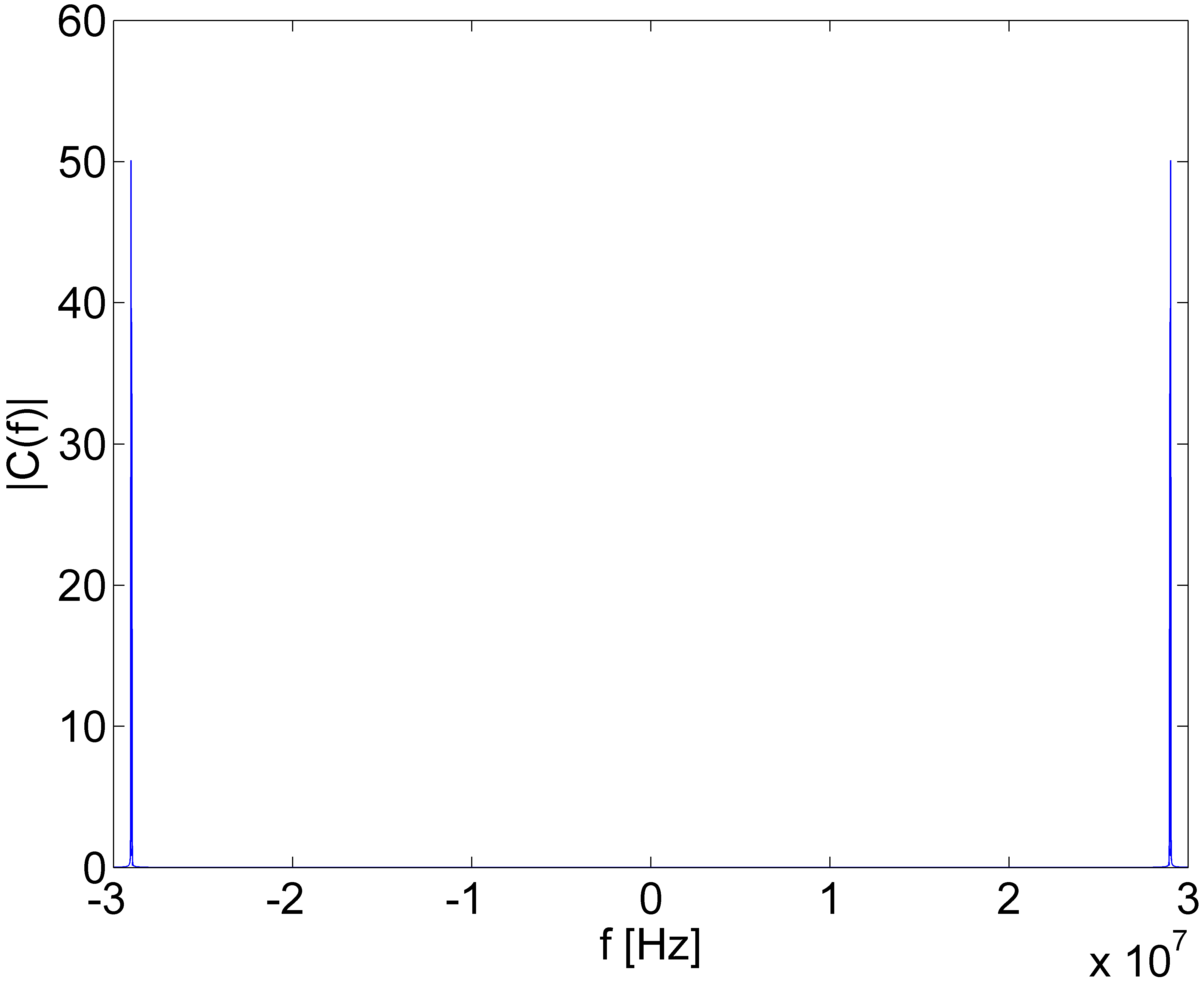}}  \\
\subfloat[]{\includegraphics[width=0.35\textwidth]{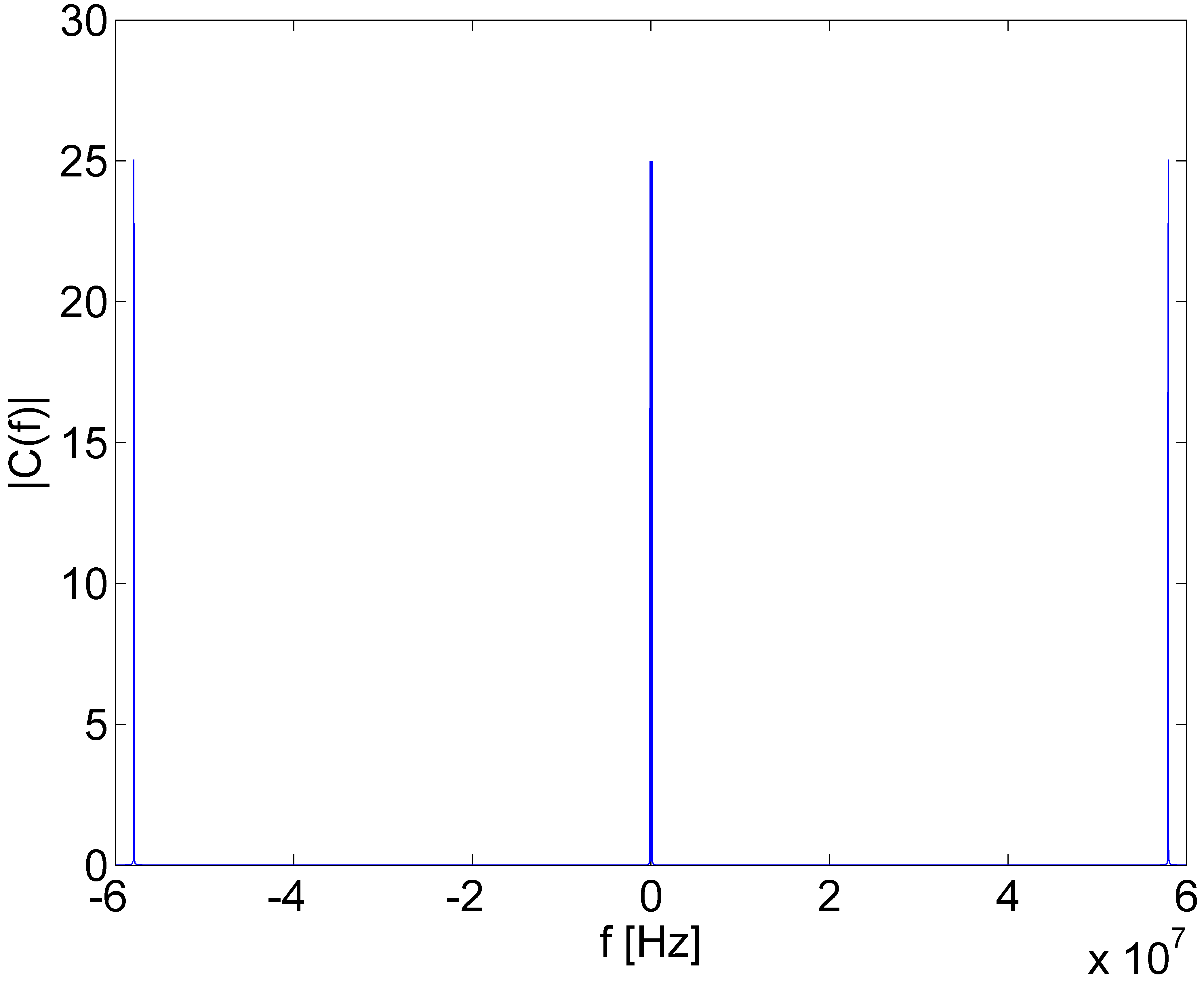}}  \subfloat[]{\includegraphics[width=0.35\textwidth]{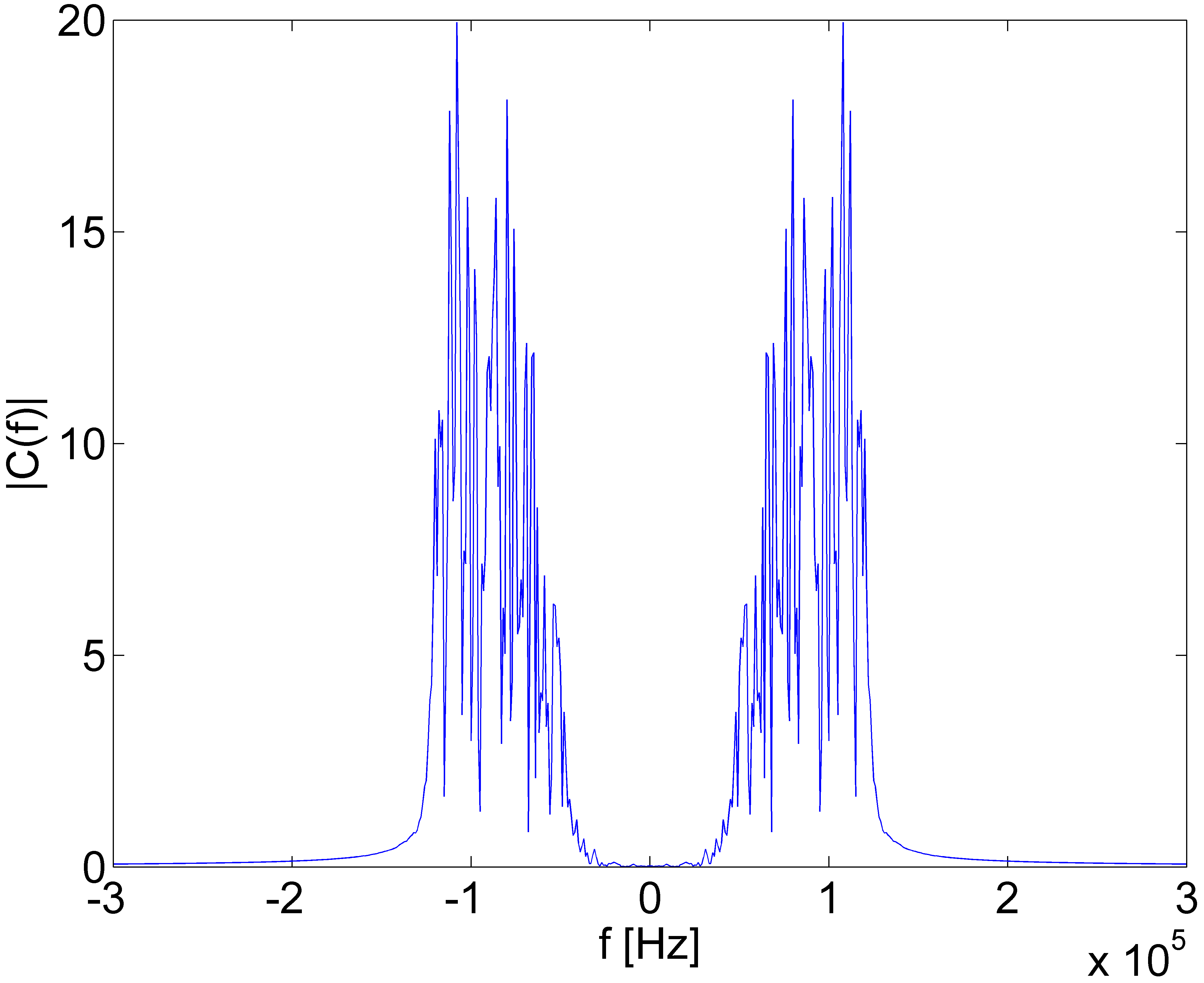}}
\caption{The dynamics of the signal's spectrum through one channel of our 4-channel crystal receiver.  (a) The signal is modulated to the crystal's center-frequency. (b) The signal's spectrum is filtered with the narrowband crystal-filter. (c) The signal is demodulated to baseband, while high image frequencies appear. (d) The signal is filtered with an anti-aliasing filter to suppress image frequencies and unaccounted nonlinear effects.}
\label{fig:Spectrum}\end{figure*}
\subsection{Analog Receiver Design}
A full block-diagram of our proposed receiver is depicted in Fig.~\ref{fig:Scheme}. The depicted system is intended to operate in baseband frequencies, while the receiving and demodulating of RF signals was simulated in a computer environment (see Section~\ref{sec:sims_exp}).  Amplification units were added, in order to compensate for the high losses along each channel. Among the reasons for these losses, are the four-fold signal power reduction at the splitter, the filtering of a $80\mbox{KHz}$ slice out of $10\mbox{MHz}$ bandwidth signal, and the total four-fold reduction caused by the up and down conversions.  As crystal filters operate only in specific central frequencies, we must perform two stages of modulation, in which the desired band is up-converted to the pass-band of the filter and then, after filtering, is demodulated to baseband. It is then filtered with a $125\mbox{KHz}$ LPF that serves to suppress the modulation image, as well as an anti-aliasing filter. 

The $125\mbox{KHz}$ stop frequency was selected in order to acquire an oversampling margin that allows minimal distortion to the Fourier coefficients lying in the $25\mbox{KHz}$-$105\mbox{KHz}$ band due to image frequencies or aliasing. 

The four-stage process is illustrated in Fig.~\ref{fig:Spectrum}, which shows the dynamics of the signal's spectrum through one of the receiver's channels.  The final LPF allows sampling each channel at $250\mbox{KHz}$, resulting in a total sampling rate of $1\mbox{MHz}$ at all channels combined -- a 20-fold reduction relative to the pulse's $20\mbox{MHz}$ Nyquist rate, and a 30-fold reduction relative to a practical implementation of MF. Once the channel output is sampled, its spectrum is easily calculated via the FFT algorithm, and the 320 relevant Fourier coefficients are ready as the recovery algorithm's input.

In our design we make a use of 3 filtering stages. The first and second stages are located before the mixers. When using mixers in data paths, special attention must be given in order to suppress image frequencies. This is of great importance as image frequency coinciding even partially with the data Fourier coefficients results in non-invertible distortions. In the first stage modulation to the crystal pass band is performed. Since the modulated frequencies and the images vary from channel to channel the specification for the preliminary LPF must be set to the worst case. Thus, the left-most image frequency that modulates our data, which is $56.35\mbox{MHz}$, must be sufficiently suppressed by the LPF. Simulations affirmed that at a 40dB attenuation this frequency does not affect the performance, and therefore this is the minimal attenuation we require. Table~\ref{table:Pre_LPF} summarizes the specifications for the preliminary filter.

\begin{table}[ht]
\centering
\begin{tabular}{| l | l |}
\hline
 Parameter & Value \\\hline
Pass Frequency & $2.5\mbox{MHz}$  \\
Maximal Passband Ripple & $1\mbox{dB}$ \\
Stop Frequency &$56.35\mbox{MHz}$ \\
Minimal Stopband Attenuation & $40\mbox{dB}$ \\\hline
\end{tabular}
\caption{Preliminary lowpass filter specifications.}
\label{table:Pre_LPF}
\end{table}

\begin{figure}[t!]
\includegraphics[width=0.5\textwidth]{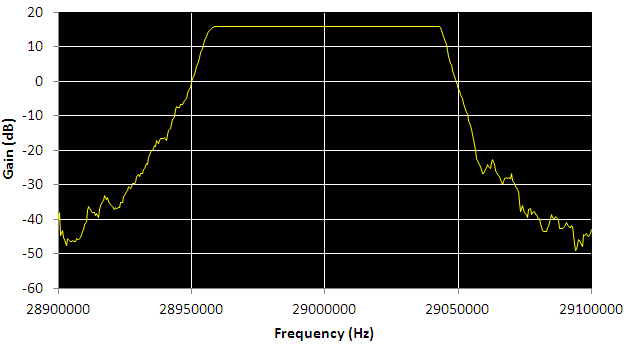}
\caption{Magnitude response of the Crystal BPF filter as a function of frequency.}
\label{fig:Crystal}
\end{figure}

In the next filtering stage, we make use of crystal filters. As these are standard, of-the-shelf, devices, the rest of the stages must be adapted to their properties, while maximizing the channel efficiency (in terms of amount of data acquired and the sampling rate at the analog-to-digital converters). Fig.~\ref{fig:Crystal} shows the magnitude response of a crystal filter, as was measured in a network-analyzer, and Table~\ref{table:Crystal} details its properties. With a data bandwidth of $80\mbox{KHz}$ on each channel, we are able to acquire as many as 320 Fourier coefficients, which allows reconstruction of various scenarios including a different number of targets, varying distances and a wide RCS range. The narrow transition band of the magnitude response, which achieves $60\mbox{dB}$ attenuation at an offset of $60\mbox{KHz}$, allows to demodulate the data band to very low frequencies. In practice, we demodulate the left-most frequency of the pass-band to a frequency of $25\mbox{KHz}$, with the right-most demodulated to $105\mbox{KHz}$. Demodulating the data band to the DC frequency is undesirable, since in this case image suppression is not satisfactory.

\begin{table}[ht]
\centering
\begin{tabular}{| l | l |}
\hline
 Parameter & Value \\\hline
Center Frequency & $29\mbox{MHz}$  \\
$-3\mbox{dB}$ Bandwidth & $80\mbox{KHz}$  \\
Maximal Passband Ripple &$1\mbox{dB}$ \\
Stopband Frequencies & $28.94\mbox{MHz}$, $29.06\mbox{MHz}$ \\
Minimal Stopband Attenuation & $60\mbox{dB}$ \\\hline
\end{tabular}
\caption{Crystal bandpass filter characteristics.}
\label{table:Crystal}
\end{table}

Finally, sampling the signal requires that we suppress any frequency that might alias our data. Examining Fig.~\ref{fig:Spectrum}(c), it is evident that the main contribution to aliasing will occur from the part of the crystal's passband energy that was modulated to very high frequencies. Therefore, it is sufficient to design an anti-aliasing filter that is efficient enough in suppressing that band. However, each channel incorporates the usage of amplifiers, which have non-linear regimes. Such non-linearities may introduce high-order harmonies of our signal, and therefore the anti-aliasing filter must be designed to suppress any frequency component which we might not have accounted for. Using narrow filters also contributes to noise reduction, which improves the performance of the recovery algorithm. The desired characteristics of the anti-aliasing LPF are detailed in Table~\ref{table:AALPF}. With the stop frequency being $125\mbox{KHz}$ we are able to sample the output signal at a rate of $250\mbox{KHz}$, which yields a total $1\mbox{MHz}$ rate at all four channels.

\begin{table}[ht]
\centering
\begin{tabular}{| l | l |}
\hline
 Parameter & Value\\\hline
Pass Frequency & $105\mbox{KHz}$  \\
Maximal Passband Ripple &$1\mbox{dB}$ \\
Stop Frequency & $125\mbox{KHz}$ \\
Minimal Stopband Attenuation & $25\mbox{dB}$ \\\hline
\end{tabular}
\caption{Anti-aliasing lowpass filter specifications.}
\label{table:AALPF}
\end{table}

Another consideration in the design of the receiver is the noise figure (NF) of the system. The NF measures the degradation of the SNR from the beginning to end of the channel. In our design, we make use of operational amplifiers LMH-6629 as power amplifiers, which have a gain of $14\mbox{dB}$ and an NF of $9\mbox{dB}$. The mixers have a conversion loss of $5\mbox{dB}$, while the crystal filters have an insertion loss of $4\mbox{dB}$. Splitting the power of the signal to four separate channels yields attenuation of $6.28\mbox{dB}$. The preliminary LPF and the anti-aliasing LPF have very small insertion loss, and therefore their NF's are negligible. To obtain the total NF of the system, we use the Friis equation for the total noise figure {\cite{Friis}},
\begin{equation}
NF_{system}=NF_{1}+\sum_{n=2}^{N}\frac{F_n-1}{\prod_{m=1}^{n-1} G_{m}}
\label{eq:Friis}
\end{equation}
where $G_n$ and $NF_{n}$ are the gain and noise figure of the $n\mbox{th}$ component in the channel, respectively, and $N$ is the number of total components. Incorporating the fact that the NF of an attenuator is equal to its attenuation level, the total NF of our system equals $NF_{system}=11\mbox{dB}$. This causes a degradation that can be compensated by averaging over several periods of the signal, as discussed in Section~\ref{sec:sims_exp}.

\begin{figure}[t!]
\centering
\includegraphics[width=0.4\textwidth]{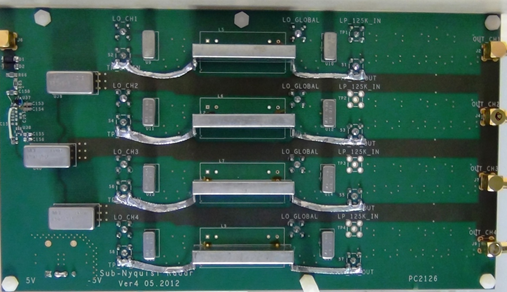}
\caption{The 4-channel crystal receiver prototype.}
\label{fig:Board}
\end{figure}

We present a photo of our 4-channel crystal receiver analog board prototype in Fig.~\ref{fig:Board}. In the next section we show real-time experiments of our hardware prototype, and compare our scheme's performance to the simple LPF channel which filters the signal at $500\mbox{KHz}$ and samples it at an identical rate of $1\mbox{MHz}$, as well as to the traditional MF which operates at rates higher than Nyquist.


\section{Simulations and Experiments}
\label{sec:sims_exp}
In this section we present hardware simulations which simulate practical filtering, as well as real experiments of our hardware prototype that include the transmission of analog signals.
\subsection{Simulations}
\begin{figure}[t!]
\centering
\subfloat[]{\includegraphics[width=0.4\textwidth]{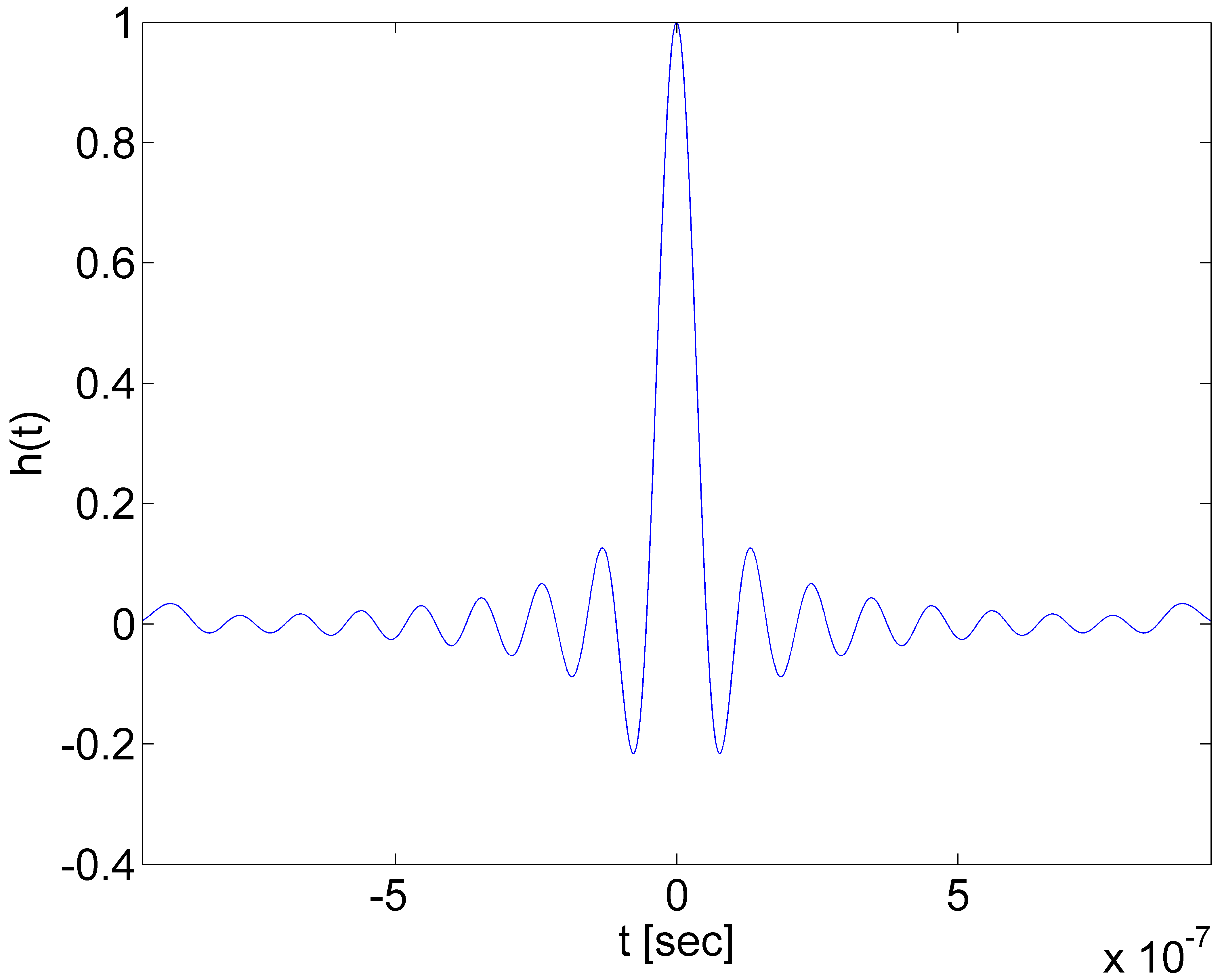}}  \\
\subfloat[]{\includegraphics[width=0.4\textwidth]{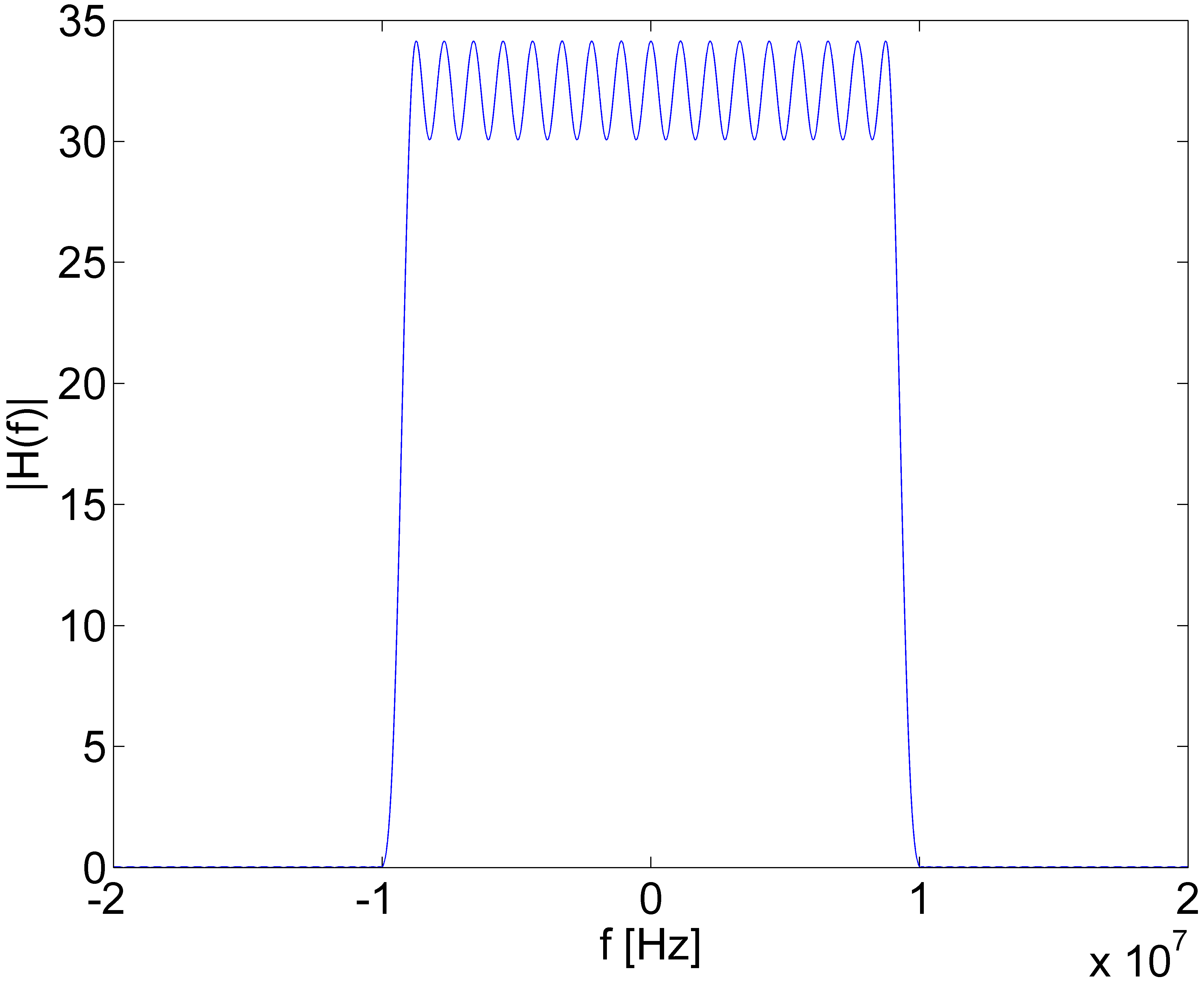}}
\caption{The temporal shape of the pulse (a) and its Fourier spectrum (b).}
\label{fig:Pulse}
\end{figure}

The first evaluation step we introduce is simulating our hardware and recovery method via MATLAB. We examined our technique's success in recovering multiple realizations of the signal defined in~\eqref{eq:Model}, after its corruption by noise. Each realization comprised $L=6$ pulses.  The pulse and its spectrum are depicted in Fig.~\ref{fig:Pulse}. Time delays and amplitudes were drawn uniformly, at random, within the intervals [0,1msec) and [0.5,1.5] respectively. The signals were generated digitally, at a rate of 0.6GHz, which is much higher than the Nyquist rate corresponding to $h(t)$'s 10MHz bandwidth. The signals were corrupted by zero-mean white Gaussian noise, with variance $\sigma^{2}$ determined such that the SNR, defined with respect to the weakest target as
\begin{equation}
\mbox{SNR}=\frac{1}{\sigma^{2}}\underset{l=1,\ldots,L}{\min}\left|a_l\right|^{2}
\label{eq:SNRDef}
\end{equation}
maintains a predefined value.

To measure the system's recovery performance and to compare it to other solutions, we define the Hit-Rate and the root-mean-square error (RMSE) metrics in the following manner:
\begin{equation}
\begin{split}
\text{Hit-Rate}&=\frac{1}{L}\left|\{ \hat{t}_l \vert \left| \hat{t}_l-t_l \right| \leq \epsilon_{th}, l=1,\ldots,L  \}\right| \\
\text{RMSE}&=\left(
      \frac{1}{L\cdot \text{Hit-Rate}}
      \sum_{l: \left| \hat{t}_l-t_l \right| \leq \epsilon_{th}}
      \left( \hat{t}_l-t_l  \right)^2
    \right)^{\frac{1}{2}},
\end{split}
\label{eq:PerfMetrics}
\end{equation}
where $\{\hat{t}_l\}_{l=1}^{L}$ are the estimated time delays and $\epsilon_{th}$ is the tolerance factor, determined by application (we chose $\epsilon_{th}=150\mbox{ns}$, which is three times the signal's Nyquist period). The RMSE is calculated only with respect to the estimations found within the tolerance interval. In order to obtain statistically stable results, each experiment was repeated 500 times.

We evaluated the performance of four different estimation methods. The first one was our 4-channel crystal receiver, where OMP was used for recovering the unknown signal parameters from its subset of Fourier series coefficients. The next two methods were based on a single LPF channel which operates in a total sampling rate of $1\mbox{MHz}$, and obtains a group of consecutive Fourier harmonics. Although the sampling rate at this single channel is $1\mbox{MHz}$, we are only able to extract $400$ coefficients due to non-ideal transition bands. In one of the methods, a spectral analysis method, known as the matrix pencil (MP) algorithm \cite{Sarkar} was used, while in the second we used again the OMP algorithm. Note that all three of the schemes operated under the same overall sampling rate. Finally, we compared the former approaches with the traditional MF method, which requires sampling at a rate 1.5 times higher the Nyquist rate ($30\mbox{MHz}$).

All the filters used in our simulations were modeled via MATLAB's filter design and analysis tool, and were based on the Chebyshev type-I filter. The electronic filters in the different schemes (such as the anti-aliasing LPF's) were designed so that their order would not increase above $6$. In contrast, the crystal filter was modeled with a high order BPF, in order to mimic its narrow transition bands.
As part of the digital procedure, we integrated over reflections of 500 consecutive pulses in order to reduce noise power. While this procedure constrains us to slow moving targets, our technique remains applicable to various scenarios, such as naval target, vehicle and human tracking, etc.

\begin{figure}
\centering
\subfloat[]{\includegraphics[width=2.8in]{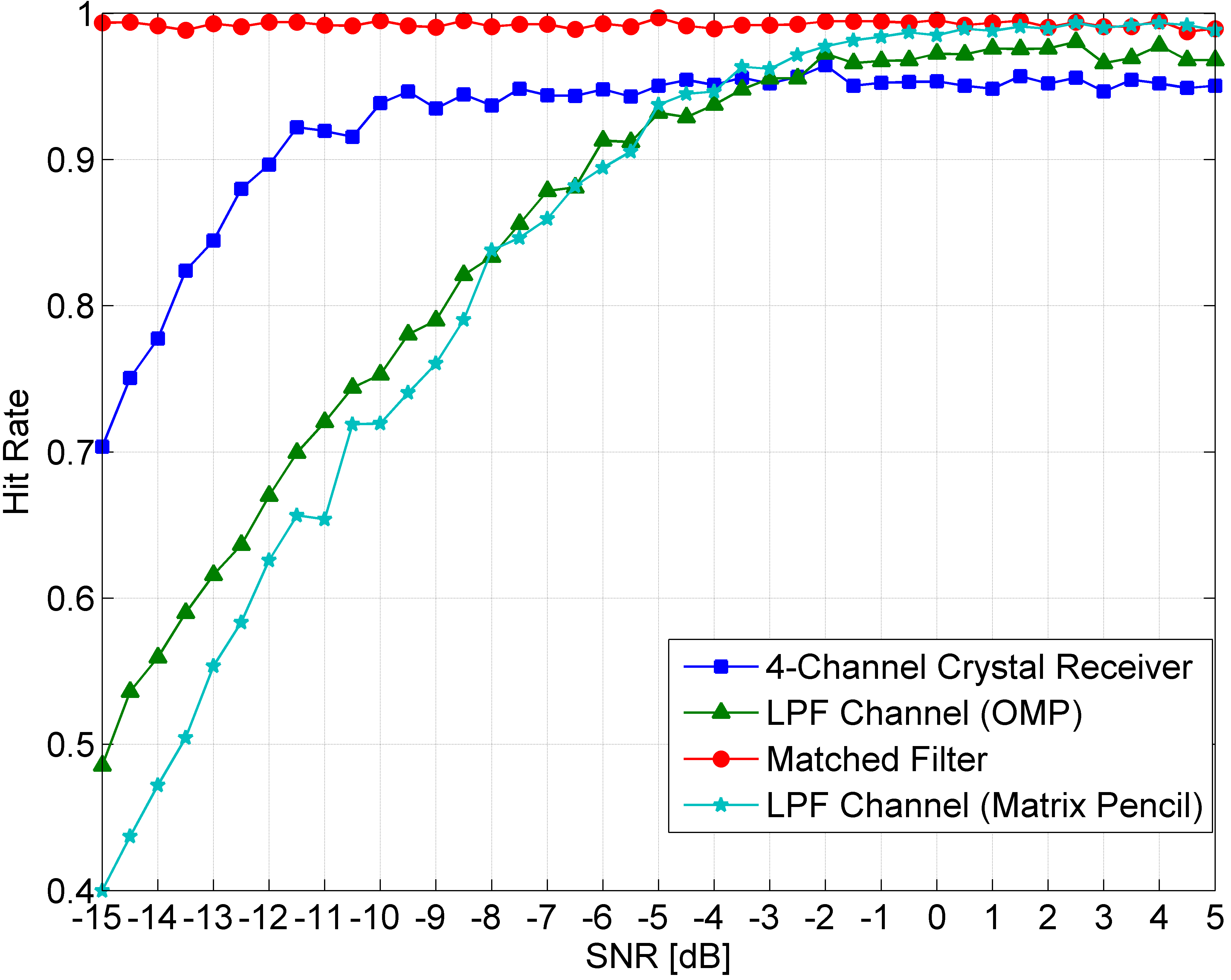}} \hfil \subfloat[]{\includegraphics[width=2.8in]{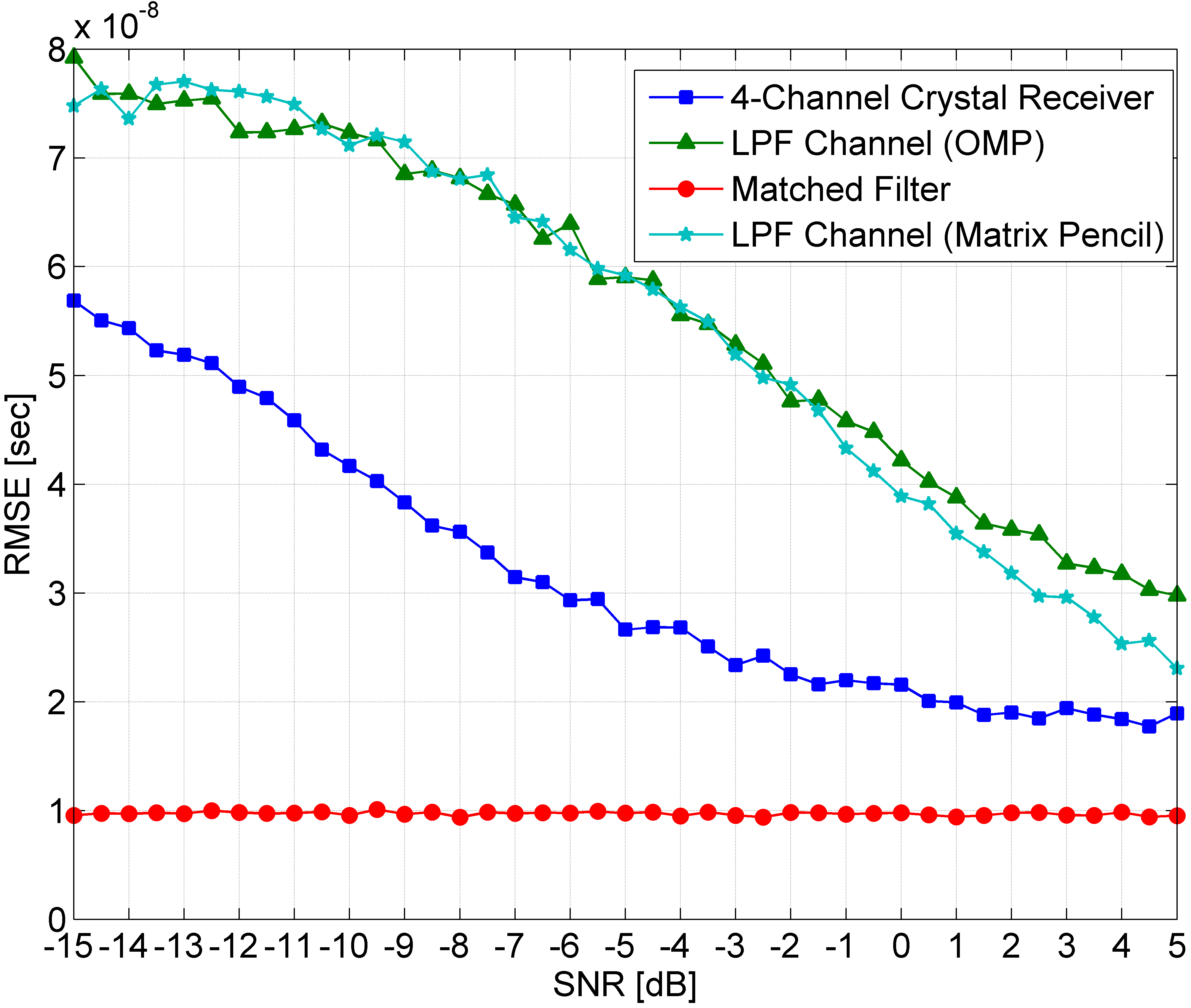}}
\caption{A comparison of four estimation methods. (a) Hit Rate (b) RMSE. The number of samples per PRI used for processing is: 1000 (Crystal Receiver, LPF and MP), 30000 (MF). }
\label{fig:ROC}
\end{figure}

Fig.~\ref{fig:ROC} depicts the performance of the four aforementioned methods, as a function of the pre-integrated SNR, tested for multiple realizations of our radar signal. Note that two of the methods make use of the same single LPF channel. Examining the results we infer that our 4-channel crystal receiver yields much better performance than LPF based methods at lower SNR values, an outcome that makes sense as in noisy realizations the aperture is critical. As the SNR increases though, the aperture plays a less significant role, and the total number of samples determines the reconstruction performance. It is not surprising that the less noisy the samples are the better the performance is obtained by MP. The latter does not quantize the time axis and is proved to reach numerical precision for high enough SNR values \cite{Sarkar}. For complete comparison, the MF curve is plotted. It is quite clear that it provides better performance than our approach, in particular for lower SNR values, but as the SNR improves the performance gap drops. We also infer that our 4-channel crystal receiver provides lower RMSE than the LPF channel based methods, in the entire SNR range, and it gets closer to the MF's performance as the SNR increases. Therefore, we conclude that our system offers the best recovery performance apart from MF (which requires fast sampling rate by Nyquist) in the SNR range that is effective for realistic radar scenarios.

\subsection{Experimental Environment}
\begin{figure}[t!]
\centering
\includegraphics[width=0.5\textwidth]{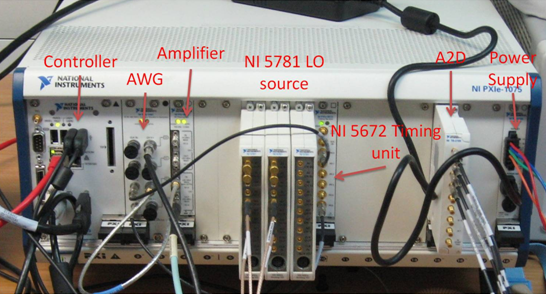}
\caption{The NI Chassis.}
\label{fig:Chassis}
\end{figure}

To evaluate the board we make use of NI PXI equipment for both system synchronization and signal sources. The entire component ensemble, wrapped in the NI chassis, is depicted in Fig.~\ref{fig:Chassis}. The components we make use of are listed in Table~\ref{table:NI_PXI}.

In order to achieve system stability, and to obtain fine recovery performance, synchronizing the system is essential. Jitters and drifts between clocks, might skew the time instances in a manner that distorts the calculated Fourier coefficients. We therefore must ensure that our entire component ensemble, including the AWG, LO signals and ADCs, are triggered by the same clock. This promises that phase unbalance between devices is brought to a minimum.

The advantage of the NI PXI chassis is that it can synchronize several devices to one base clock and distribute a trigger signal with skew less than $1\mbox{ns}$. The ability to synchronize all devices to one clock keeps the jitter between signals low so that we can assume it does not cause a change in phase during operation.  Therefore, the whole system is stable and consistent.
\begin{table}[ht]
\centering
\begin{tabularx}{0.5\textwidth}{ | l | X | }
\hline
 NI Part No. & Description  \\\hline
NI PXIe-1075 & 18-Slot 3U PXI Express Chassis  \\
NI PXIe-8133 &Core i7-820QM $1.73\mbox{GHz}$ Controller \\
NI PXIe-5451 & $400\mbox{MSamp/s}$ Arbitrary Waveform Generator (AWG) \\
NI PXI-5690 & 2 Channel RF Preamplifier from $500\mbox{KHz}$ to $3.0\mbox{GHz}$  \\
NI PXI-4130 & Power SMU \\
NI PXI-6123 & 16-Bit, $500\mbox{KSamp/s/ch}$, Simultaneous Sampling Multifunction ADC \\
NI PXIe-6672 & Timing and Multichassis Synchronization Unit for PXI Express  \\
NI 5781 &Baseband Transceiver for NI FlexRIO \\
NI PXI-7965R & NI FlexRIO FPGA Module for PXI \\\hline
\end{tabularx}
\caption{NI PXI devices.}
\label{table:NI_PXI}
\end{table}
\subsection{Experiments}
\begin{figure}[t!]
\centering
\includegraphics[width=0.5\textwidth]{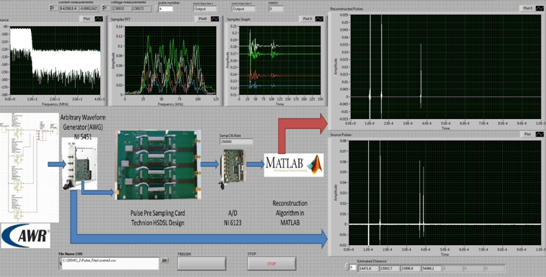}
\caption{The LabView experimental interface. From left to right: the signal's spectrum, the frequency response of each channel, the 4 signals view in each channel, at the top- the reconstructed signal, at the bottom- the transmitted signal. }
\label{fig:LabView}
\end{figure}

The experimental process consists of the following steps. We begin by using the AWR software, which gives us the ability to examine a large variety of scenarios, comprised of different targets, distances and RCS values. It is able to simulate the complete radar scenario, including the pulse transmission and accurate power loss due to wave propagation in a realistic medium. It also takes into account the reflections from the unknown targets, which are proportional to the their RCS. The AWR software also contains a model of a realistic RF receiver, which performs signal processing in high frequencies. The demodulation of the signal to IF frequencies is simulated, and the output is saved to a file.  Next, the simulation result is loaded to the AWG module, which produces an analog signal. This signal is amplified using the NI 5690 LNA and then routed to our 4-channel crystal receiver. The receiver is fed by 5 LO, of which 4 modulate the desired frequency band of each channel individually to the crystal pass band, and a global one that modulates the latter to a low frequency band, before sampling. The LO's are created using 3 NI 5781 baseband transceivers, acting as trigger based signal generators with a constant and known phase, controlled by NI Flex Rio FPGA's. The AWG also triggers the ADC to sample 250 samples in each sampling cycle, per channel. These samples are fed into the chassis' controller and a MATLAB code is launched that runs the reconstruction algorithm. Our system contains a fully detailed interface implemented in the LabView environment, which allows simple activation of the process. A screenshot of the interface is depicted in Fig.~\ref{fig:LabView}.

Performing the previously described processes, we tested various scenarios on the board - that is, a variety of targets, distances and RCS - and examined the reconstruction quality. We tested our board for its limitations and have obtained the following results. The minimal voltage for a reconstructible pulse, without PRI integration, is approximately $4\mbox{mV}$, which corresponds to a power of $P_{min}=320\mbox{nW}=-34.94\mbox{dBm}$. As the board contains amplification units, we must also consider the maximal allowable input power. Above this power, the amplifiers reach their non-linear regimes and introduce non-invertible distortions to the input signal. It was measured that above a voltage of $92\mbox{mV}$, which corresponds to a power of $P_{max}=169.28\mu\mbox{W}=-7.7139\mbox{dBm}$, such effects begin to take place. We thus deduce that our system has a dynamic range (DR) of
\begin{equation}
DR_{system}=P_{max}-P_{min}=27.23\mbox{dB}.
\label{eq:DR}
\end{equation}
The DR of the analog system can be digitally enhanced using pulse integration. For example, integrating over $500$ pulses, increases the DR by a factor of $10\log_{10}(500)=27\mbox{dB}$.

Moreover, we have tested the limitations for multiple-pulse scenario. In cases where the OMP algorithm fails to obtain time delays with a small enough error, calculating the residual might yield unsatisfactory attenuation of the part of the measurements corresponding to the located pulse. For pulses energetic enough this means that their residual might still be stronger than the remaining pulses. We have measured that a scenario of two pulses of which voltage ratio is above $12$, causes the algorithm to select the stronger pulse twice in consecutive iterations, thus failing to identify the weaker pulse.

\begin{figure}
\centering
\subfloat[]{\includegraphics[width=0.5\textwidth]{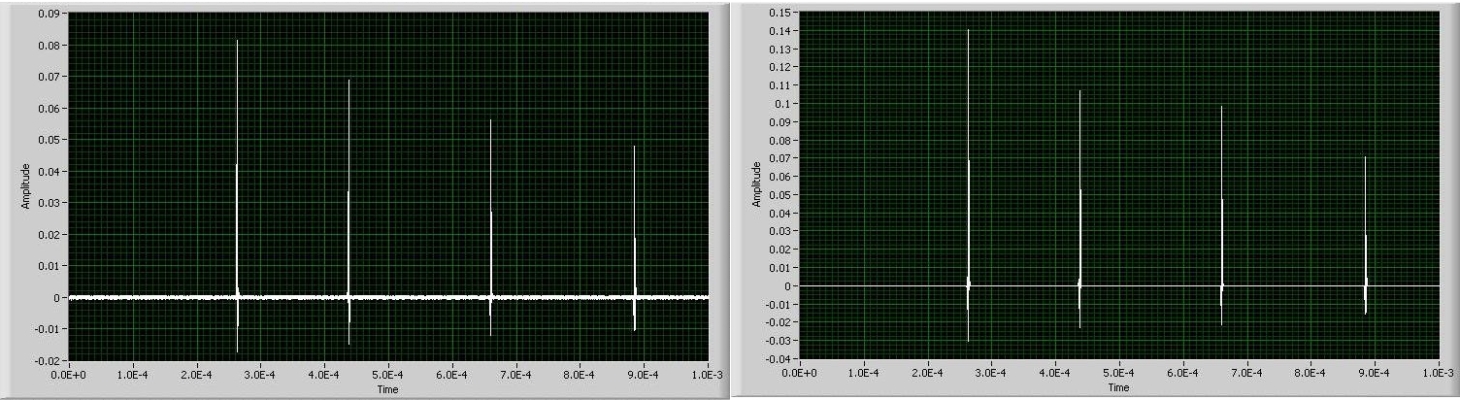}} \hfil
\subfloat[]{\includegraphics[width=0.5\textwidth]{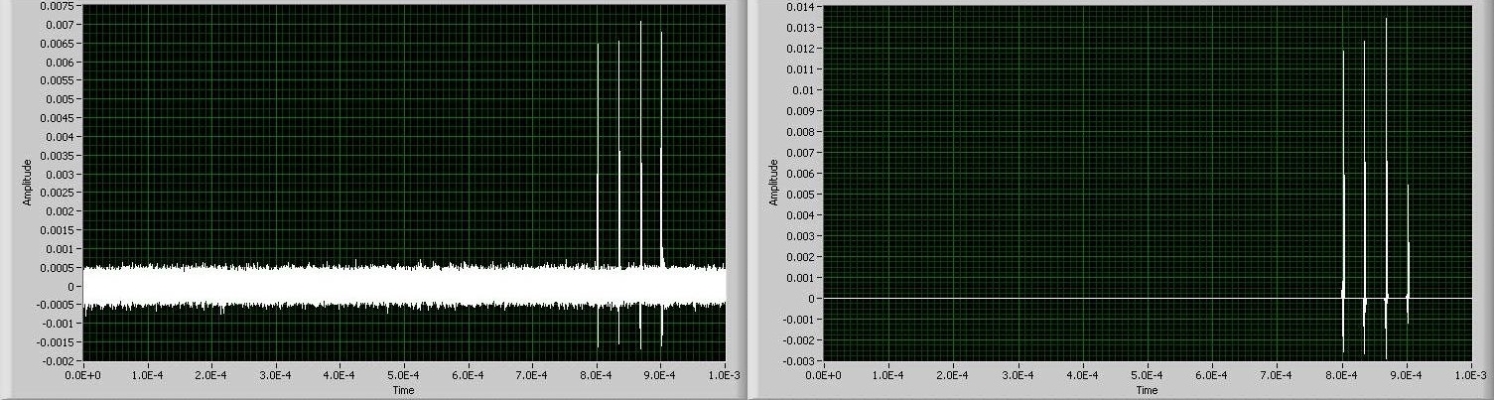}} \hfil
 \subfloat[]{\includegraphics[width=0.5\textwidth]{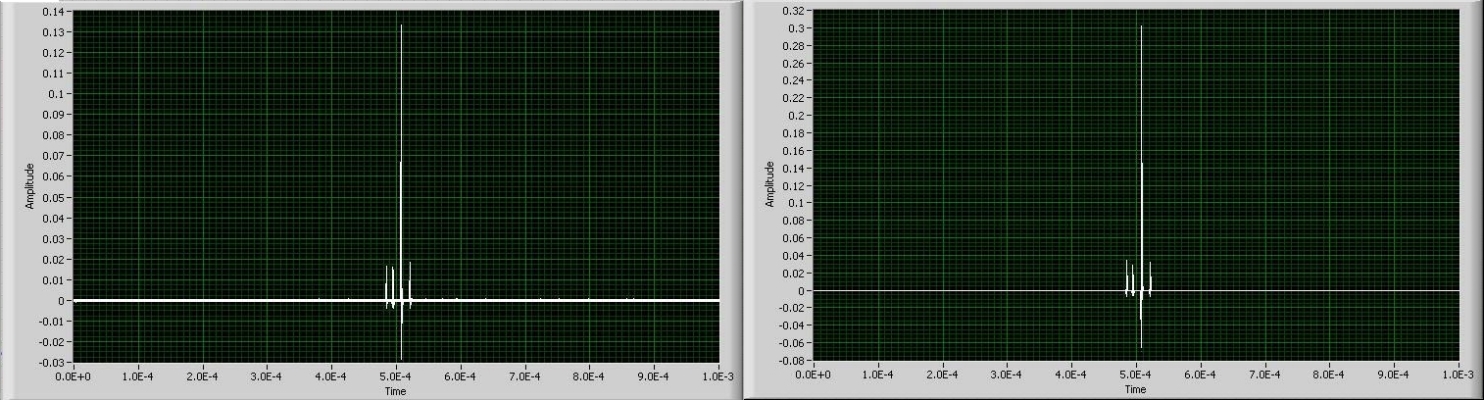}}
\caption{Signal reconstruction with the 4-channel crystal receiver, for different scenarios. On the left are the received signals, while on the right are the reconstructed signals.}
\label{fig:New_Exp}
\end{figure}

Fig.~\ref{fig:New_Exp} demonstrates our system's reconstrution abilities in 4 target scenarios. In (a), a scenario of 4 evenly spaced targets was simulated. The targets were located at the following distances- 39.28km; 65.43km; 98.76km; 132.54km. In this case, the provided estimation was 39.275km; 65.423; 98.752km; 132.529km, respectively, providing a maximal error of 11m, which is indeed quite impressive. (b) tests evenly spaced targets at shorter distances. The received pulses correspond to targets at distances 120km; 125km; 130km; 135km. Note that this experiment shows the system's performance in relatively long distances. Estimated targets were acquired at 119.99km; 124.995km; 129.997km; 134.981km, providing a maximal error of 19m. In (c) we demonstrate the system's dynamic range, as well as its ability to separate close targets. The transmitted pulses correspond to targets located at 72.5km; 74km; 76km; 78km. To demonstrate the dynamic range the third target was chosen to have an echo 7 times stronger than the other targets. The system estimtated the following distances- 72.499km; 73.992km; 75.994km; 77.995km. Furthermore, although the pulses' amplitudes were highly diverse, their estimations were accurate.

\section{Conclusions}
\label{sec:Conclusions}
We presented the first sub-Nyquist radar prototype based on the Xampling methodology. We have been able to reduce the total sampling rate by a factor of approximately 30 while maintaining reasonable target location estimation and hit rate ability.

Our experiments with the hardware have further proven that sub-Nyquist sampling of radar signals is possible without losing much of the recovery performance.

In the future, theoretical research might enrich the mathematical model, with the addition of Doppler effect and relaxing the demand that the shape of the pulses is known. Different algorithms and methods might be considered instead of OMP, so that noise robustness and spatial resolution might increase. However, we believe that the actual implementation of the sub-Nyquist radar prototype, is an important and crucial step towards the incorporation of Xampling and FRI frameworks into real communications and signal-processing systems.

\section*{Acknowledgments}

The authors would like to thank National Instruments corporation for their support throughout the development of the prototype, and for their supply of equipment that allowed system operation. Special acknowledgments are due to Mr. Eran Castiel of NI Israel, Dr. Ahsan Aziz of NI Texas, and the entire support group of NI Israel, for their efforts and full cooperation with the sub-Nyquist group at the Technion.

The authors would also like to thank Dr. Eyal Doron of “Pearls of Wisdom, Advanced Technologies”, Kefar Netter, Israel, for sharing valuable insights from the field of array processing and for many helpful discussions.

\bibliographystyle{IEEEbib}
\bibliography{citations}
\end{document}